\documentclass[apj]{emulateapj}

\shorttitle{Temperatures and line formation in fan loops}        
\shortauthors{Brooks et al.}

\begin{document}

\title{EUV spectral line formation and the temperature structure of active region fan loops: observations with {\it Hinode}/EIS and {\it SDO}/AIA}
\author{David H. Brooks \altaffilmark{1,2}, Harry P. Warren \altaffilmark{3}, Peter R. Young\altaffilmark{1}}
\altaffiltext{1}{College of Science, George Mason University, 4400 University Drive, Fairfax, VA 22020}                           
\altaffiltext{2}{Present address: Hinode Team, ISAS/JAXA, 3-1-1 Yoshinodai, Sagamihara, Kanagawa 229-8510, Japan}
\altaffiltext{3}{Space Science Division, Naval Research Laboratory, Washington, DC 20375}
\email{dhbrooks@ssd5.nrl.navy.mil}

\begin{abstract}
With the aim of studying active region fan loops 
using observations from the {\it Hinode} EUV Imaging Spectrometer (EIS) and
{\it Solar Dynamics Observatory} (SDO) Atmospheric Imaging Assembly (AIA), we investigate
a number of inconsistencies in modeling the absolute intensities of \ion{Fe}{8} and \ion{Si}{7} lines, 
and address why spectroheliograms formed from these lines look very similar
despite the fact that ionization equilibrium calculations suggest that they have significantly different 
formation temperatures: $\log\,(T_e/K)$ = 5.6 and 5.8, respectively. These issues 
are important to resolve because confidence has been 
undermined in their use for differential emission measure (DEM) analysis, and \ion{Fe}{8}
is the main contributor to the AIA 131\,\AA\, channel at low temperatures. Furthermore, 
the strong \ion{Fe}{8} 185.213\,\AA\, and \ion{Si}{7}
275.368\,\AA\, lines 
are the best EIS lines to use for velocity studies in the transition region, and 
for assigning the correct
temperature to velocity measurements in the fans. 
We find that 
the \ion{Fe}{8} 185.213\,\AA\, line is
particularly sensitive to the slope of the DEM, leading to  
disproportionate changes in its effective
formation temperature. If the DEM has a steep gradient 
in the $\log\,(T_e/K)$ = 5.6 to 5.8 temperature range, or is strongly peaked, \ion{Fe}{8} 185.213\,\AA\, and \ion{Si}{7}
275.368\,\AA\, will be formed at the same temperature. We show that this effect explains the similarity
of these images in the fans. Furthermore, 
we show that the most recent ionization balance compilations resolve the 
discrepancies in absolute intensities. With these difficulties overcome, we combine EIS and AIA data to 
determine the temperature structure of a number of fan loops and find that they have peak temperatures of 0.8--1.2MK.
The EIS data indicate that the temperature distribution has a finite (but narrow) width $<$ $\log\, (\sigma_{T_e}/K)$ = 5.5 which,
in one detailed case, 
is found to broaden substantially towards the loop base. 
AIA and EIS yield similar results on the temperature, emission measure magnitude, and thermal distribution
in the fans, though sometimes the AIA data suggest a relatively larger thermal width. The result is 
that both the \ion{Fe}{8} 185.213\,\AA\, and \ion{Si}{7}
275.368\,\AA\, lines are formed at $\log\, (T_e/K) \sim$ 5.9 in the fans, and 
the AIA 131\,\AA\, response also shifts to this temperature.
\end{abstract}
\keywords{Sun: corona---Sun: UV radiation---Techniques: spectroscopic}

\section{Introduction}
To understand how the solar corona is heated to high temperatures, it is important to explain
the heating of closed field structures such as active region loops. There have been extensive
studies of these structures, and recent progress and outstanding issues have been reviewed by \citet{klimchuk_2006}
and \citet{reale_2010}. A key diagnostic of the heating of coronal loops is the differential 
emission measure (DEM) distribution, because many coronal heating models make specific predictions
as to its form and shape. For example, nanoflare reconnection models \citep{parker_1983,parker_1988}
predict the presence of
a weak high temperature component in the DEM \citep{cargill_1995}. There have been several recent studies
that have tried to detect this emission \citep{schmelz_etal2009a,reale_etal2009a,testa_etal2010}. 

The DEM gradient, or proportion of hot and cool material, also sets constraints on impulsive, steady
or quasi-steady loop heating models \citep{warren_etal2010c,tripathi_etal2010}, and
the nanoflare model also predicts a spread in the temperature distribution within a loop due to the
incoherent heating and cooling of unresolved threads. There has been considerable debate
as to whether loops have a multi-thermal temperature distribution because of conflicting measurements
by different
instruments \citep{lenz_etal1999,aschwanden_etal1999,schmelz_etal2001}. 
The discussion hinges on issues such as background
subtraction \citep{delzanna&mason_2003}, methodology \citep{aschwanden_2002}, 
or the spatial resolution \citep{aschwanden_etal2008}, or temperature resolution
of the instruments used \citep{martens_etal2002}. Recent
observations by the {\it Hinode} \citep{kosugi_etal2007} EUV Imaging Spectrometer \citep[][EIS]{culhane_etal2007b}
suggest that loops formed near 1MK have a narrow temperature distribution, but are not isothermal 
\citep{warren_etal2008a}.

Doppler velocity measurements are another important diagnostic of the heating process, and again
there have been numerous recent studies of flows in active region loops using EIS data
\citep{doschek_etal2007,hara_etal2008,delzanna_2008,brooks&warren_2009}.
Some of the signatures of coronal heating models are expected to be quite subtle, however. The nanoflare
reconnection model predicts weak downflows at `warm' temperatures \citep{patsourakos&klimchuk_2006}, and short-lived faint
upflows at high temperatures \citep{patsourakos&klimchuk_2009}. The signatures of nanoflare heating in spectral
line profiles therefore depend sensitively on the temperature of the flows, and this also implies that accurate
measurements will allow inference of the properties of the energy release. 
It is important therefore to assign the flow temperatures as accurately as possible.

\begin{figure*}
\centering
\includegraphics[width=0.95\linewidth]{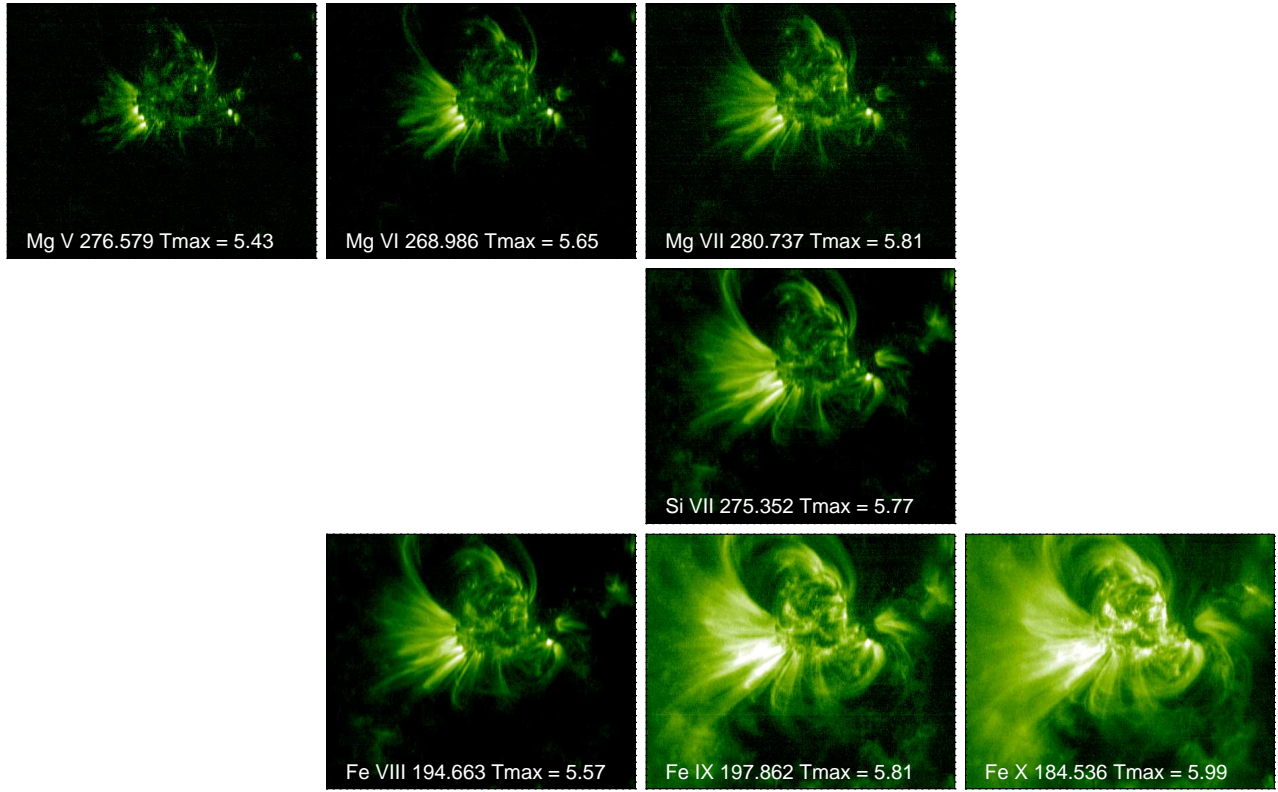}
\caption{{\it Hinode} EIS rasters of AR 10978 showing the similarity of  
\ion{Fe}{8} and \ion{Si}{7} images. The \ion{Mg}{6} and \ion{Fe}{9}
images look to be formed at lower and higher temperatures, respectively. For example, note the 
increasing vertical extent of the fan structures and cool features to east and west of the active
region as you look at the images from top to bottom of the third column (\ion{Mg}{6}, \ion{Si}{7}, \ion{Fe}{9}). 
The images are scaled linearly.
\label{fig1}}
\end{figure*}

A class of loops that have not yet been studied in sufficient detail with the latest instrumentation
are the fan structures that appear as partially
observed long loops at
the edges of active regions. They are seen mostly in the 0.4--1.3MK temperature range 
\citep{schrijver_etal1999,delzanna&mason_2003,ugarteurra_etal2009} and have densities 
greater than $\log\, (N_e/cm^{-3})$ = 9 \citep{delzanna_2003,young_etal2007b}. They also appear to show red-shifted downflows
\citep{winebarger_etal2002,marsch_etal2004}. We examine a sample of these
structures in this paper.

For this and all the above observational studies,  
spectral emission from ions of Fe is of major importance. The high elemental abundance
of Fe leads to many strong emission lines 
in spectrometer and imager pass-bands, yet 
interpretation of the observations is often difficult because of uncertainties in the atomic data
used in prediction of emission \citep{lanzafame_etal2002,young_etal2009}. 
In principle, the emission from spectral lines of Fe can provide
stringent constraints on temperatures and densities in the corona, however, it is of paramount
importance that the diagnostic capabilities of Fe lines be assessed critically.

Initial results from EIS have indicated a number of problems in interpreting the Fe emission. \citet{young_etal2007b}
noted that spectroheliograms of \ion{Fe}{8} and \ion{Si}{7} look very similar in specific active region features such as the fan loops, 
despite the fact that the temperatures of 
the peak fractional abundance in ionization 
equilibrium are significantly different; $\log\, (T_e/K)$ = 5.6 and 5.8, respectively, according to \citet{mazzotta_etal1998}.
Since Fe is a considerably more complex atom than Si and is therefore likely to be less well understood, this
observation has led to the suggestion that the ionization balance calculations for \ion{Fe}{8} need to be revised upward
to higher temperatures \citep{young_etal2007b}. Furthermore, this could have an impact on surrounding ions such as \ion{Fe}{7} and \ion{Fe}{9}. Recently, \citet{young&landi_2009} found evidence that this is indeed the case for \ion{Fe}{7}. 

In addition to this temperature problem, a number of previous differential emission measure (DEM) studies have found difficulties reproducing the
absolute intensities of \ion{Fe}{8} and \ion{Si}{7} lines simultaneously. See, for example, the quiet Sun off-limb DEM analysis of
\citet{warren&brooks_2009}, the on disk quiet Sun study of \citet{brooks_etal2009}, or the analysis of a cool active region
feature by \citet{landi&young_2009}. 

Such issues are important to resolve as the \ion{Fe}{8} 185.213\,\AA\, and \ion{Si}{7} 275.368\,\AA\, lines
would provide the best EIS lower temperature constraints on the DEM if we had good confidence
in the atomic data. Also, since they are strong, they are the best lines to use for velocity measurements in the 
transition region, and are present in the majority of EIS observations. As discussed, it is therefore of 
critical importance that 
we have confidence in the temperatures we assign to the measured velocities.

Changes to the formation temperatures or atomic data for these ions could also have 
an impact on interpreting the results from imagers. For example, changes
to \ion{Fe}{8} could affect the response functions for the 
131\,\AA\, channel of the Atmospheric Imaging Assembly (AIA) on 
the {\it Solar Dynamics Observatory (SDO)}. 

Motivated by our interest in studying the temperature and velocity structure of the fan loops, this
situation has led us to take a closer look at the formation of the \ion{Fe}{8} and \ion{Si}{7} spectral lines.
Recently, there have been a number of revisions to the ionization balance calculations \citep{bryans_etal2009,dere_etal2009}
and we investigate whether they could help resolve these inconsistencies. We also examine the details of
spectral line formation in this temperature
range in the quiet Sun and in the fan loops. In particular, we examine whether convolving the contribution
functions with the temperature distribution of the feature could explain the similarity of \ion{Fe}{8} and \ion{Si}{7} images. It is known
that the temperature of peak contribution to the line intensity can be shifted from the theoretical peak temperature
of the emissivity if the shape of the DEM is taken into consideration \citep{brosius_etal1996,feldman_etal1999,delzanna_etal2003}.
In doing so, we 
finally show that 
the ionization equilibrium calculations for Fe may not be the source of this problem. By determining more realistic
effective formation temperatures, we present
a possible explanation for the observations. We also show that the most recent ionization balance calculations can
resolve the discrepancies in the magnitudes of the intensities found in previous studies. 

With confidence in the atomic data for these ions restored, we perform an emission measure (EM) analysis of a number of 
fan loops using EIS and AIA and compare the results from the two instruments.

\section{Similarity of \ion{Fe}{8} and \ion{Si}{7} images}

Figure \ref{fig1} shows example images of AR 10978 in (top left to bottom right)
\ion{Mg}{5} 276.579\,\AA, \ion{Mg}{6} 268.986\,\AA, \ion{Mg}{7} 280.737\,\AA,
\ion{Si}{7} 275.368\,\AA, 
\ion{Fe}{8} 194.663\,\AA, \ion{Fe}{9} 197.862\,\AA, and \ion{Fe}{10} 184.536\,\AA.
These images are formed at $\log\, (T_e/K)$ = 5.4, 5.6, 5.8, 5.8, 5.6, 5.8, and 6.0, respectively, 
according to \citet{mazzotta_etal1998}.
The data were obtained on 2007, December 10, at 00:19:27UT. The EIS instrument has 4 slit
options (1$''$, 2$''$, 40$''$, and 266$''$). The 1$''$ slit was used for these observations and stepped over a 
FOV of 460$''$ by 384$''$. The exposure time was 40s at each position.
The data were 
processed using standard EIS data reduction routines available in SolarSoft (eis\_prep). 

It is clear that the images do not follow the expected temperature trend. 
The \ion{Fe}{8} and \ion{Si}{7} images look very similar, especially in the fan structures to
solar east of the AR and the cool loops to solar west. This suggests 
a similar temperature of formation. In addition, the 
\ion{Mg}{6} and \ion{Fe}{9} images look to be formed at lower and higher temperatures, respectively. 
Note, for example, the third column of the figure. The vertical extent of the cool features to solar west
and the fan structures to solar east appears to increase from top to bottom (\ion{Mg}{6} to \ion{Si}{7} to
\ion{Fe}{9}) suggesting an increase in temperatures. 
The problem identified by \citet{young_etal2007b} is clearly seen in these examples. 

\section{Atomic Data}
In this work we use several sources of atomic data as follows.
All the contribution functions are 
calculated using the CHIANTI v6.0.1 database \citep{dere_etal1997,dere_etal2009}.
The coronal abundances of \citet{feldman_etal1992} were used, together with the ionization fractions
of \citet{mazzotta_etal1998} and \citet{dere_etal2009}. 

We use four specific lines to explore how their formation temperatures change in the quiet Sun:
\ion{Mg}{6} 268.986\,\AA, \ion{Fe}{8} 185.213\,\AA, \ion{Si}{7} 275.368\,\AA, and
\ion{Fe}{9} 188.485\,\AA. For these lines we used the effective collision strengths and radiative
data of 
\citet{landi&bhatia_2007}, 
\citet{griffin_etal2000}, 
\citet{bhatia&landi_2003b}, and
\citet{storey_etal2002}, respectively. 
A large number of additional spectral lines were used 
to compute DEMs and the data sources for these secondary data are too numerous to list here.
We refer the interested reader to the CHIANTI database where the references are given in detail. 

\section{Spectral Line Formation Temperatures}
\label{slft}
\begin{figure}[ht]
\centering
\includegraphics[width=0.95\linewidth]{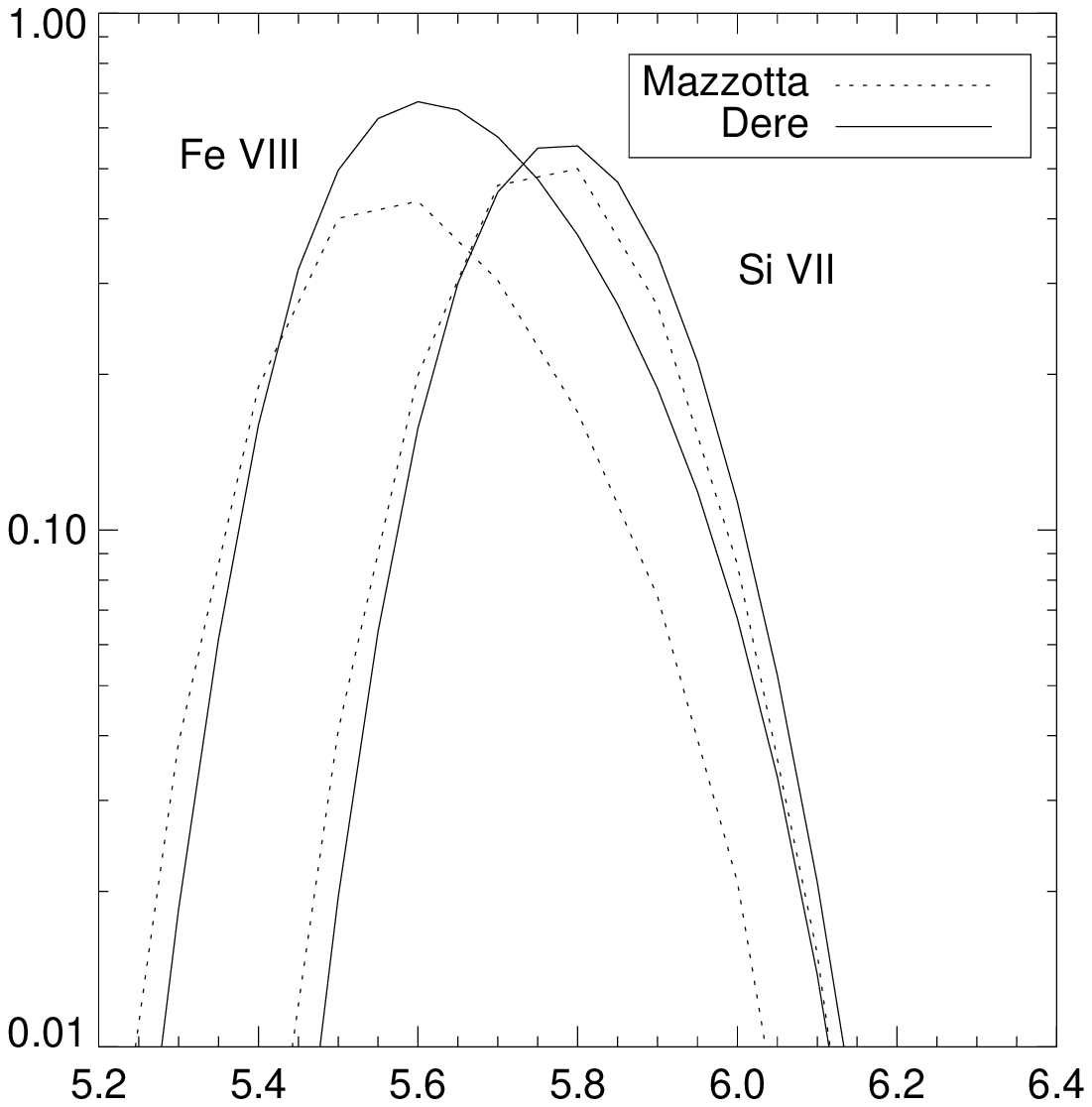}
\caption{Comparison of the \ion{Fe}{8} and \ion{Si}{7} ionization equilibrium curves calculated by \citet{mazzotta_etal1998} with those
of \citet{dere_etal2009}.
\label{fig2}}
\end{figure}

The intensity of an optically thin emission line arising from an atomic transition from level $i$ to
level $j$ can be expressed as
\begin{equation}
I_{ij} = \int_{T_e} G_{ij} (T_e, N_e) \phi (T_e) dT_e 
\label{eq1} 
\end{equation}
where 
the differential emission measure (DEM) is defined as, 
\begin{equation}
\phi (T_e) = N_e^2 dz/dT_e
\end{equation}
with $dz$ the differential distance along the line of sight 
and $N_e$ the electron density. The contribution function is defined as
\begin{equation}
G_{ij} (T_e, N_e) = A(Z) F (T_e, N_e) \epsilon_{ij} (T_e, N_e) { n_H \over (4 \pi N_e^2) }
\label{eq2}
\end{equation}
with $A(Z)$ the elemental abundance, $n_H$ the hydrogen number density, $\epsilon_{ij} (T_e, N_e)$
the line emissivity normalized to the ground state population, and $F (T_e, N_e)$ the
ionization fraction as a function of temperature and density. The accuracy of the computation of
$F$ for Fe and Si is important for this paper. In Figure \ref{fig2} we contrast the \citet{mazzotta_etal1998}
ionization equilbrium calculations for \ion{Fe}{8} and \ion{Si}{7} with the updated results of \citet{dere_etal2009}.
Note that the peak formation temperature does not change, so the temperature problem remains. 
The maximum fractional
abundance, however, increases by 10\% for \ion{Si}{7} and 55\% for \ion{Fe}{8}. These changes could
be important when modeling the magnitudes of the absolute intensities. Note that the 
\citet{bryans_etal2009} calculations give the same formation temperatures for \ion{Si}{7} and \ion{Fe}{8} as
\citet{dere_etal2009}. 
The maximum fractional abundances are also within 10\%.

\begin{figure*}[t!]
\centering
\includegraphics[width=1.00\linewidth]{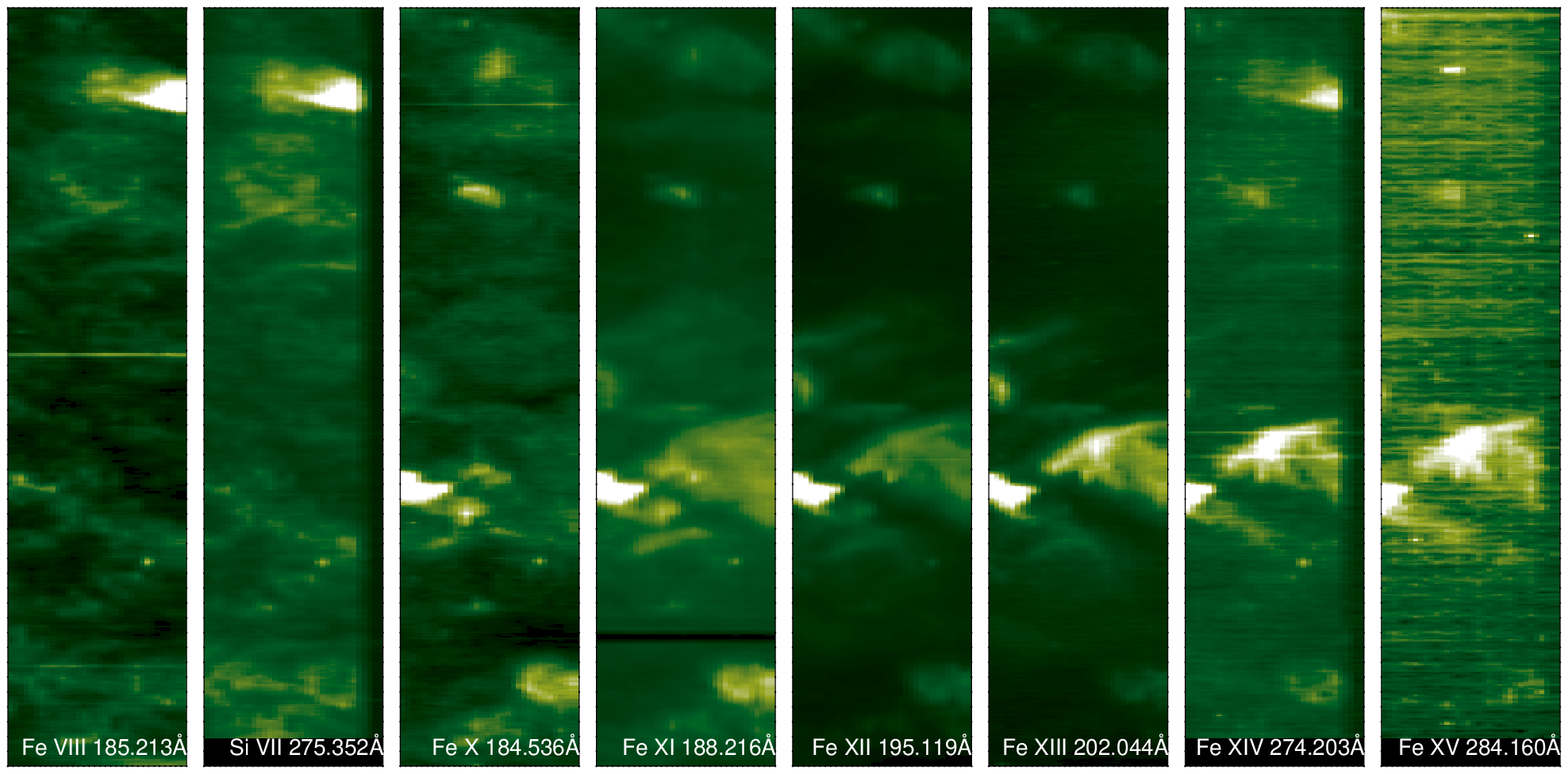}
\includegraphics[width=0.45\linewidth]{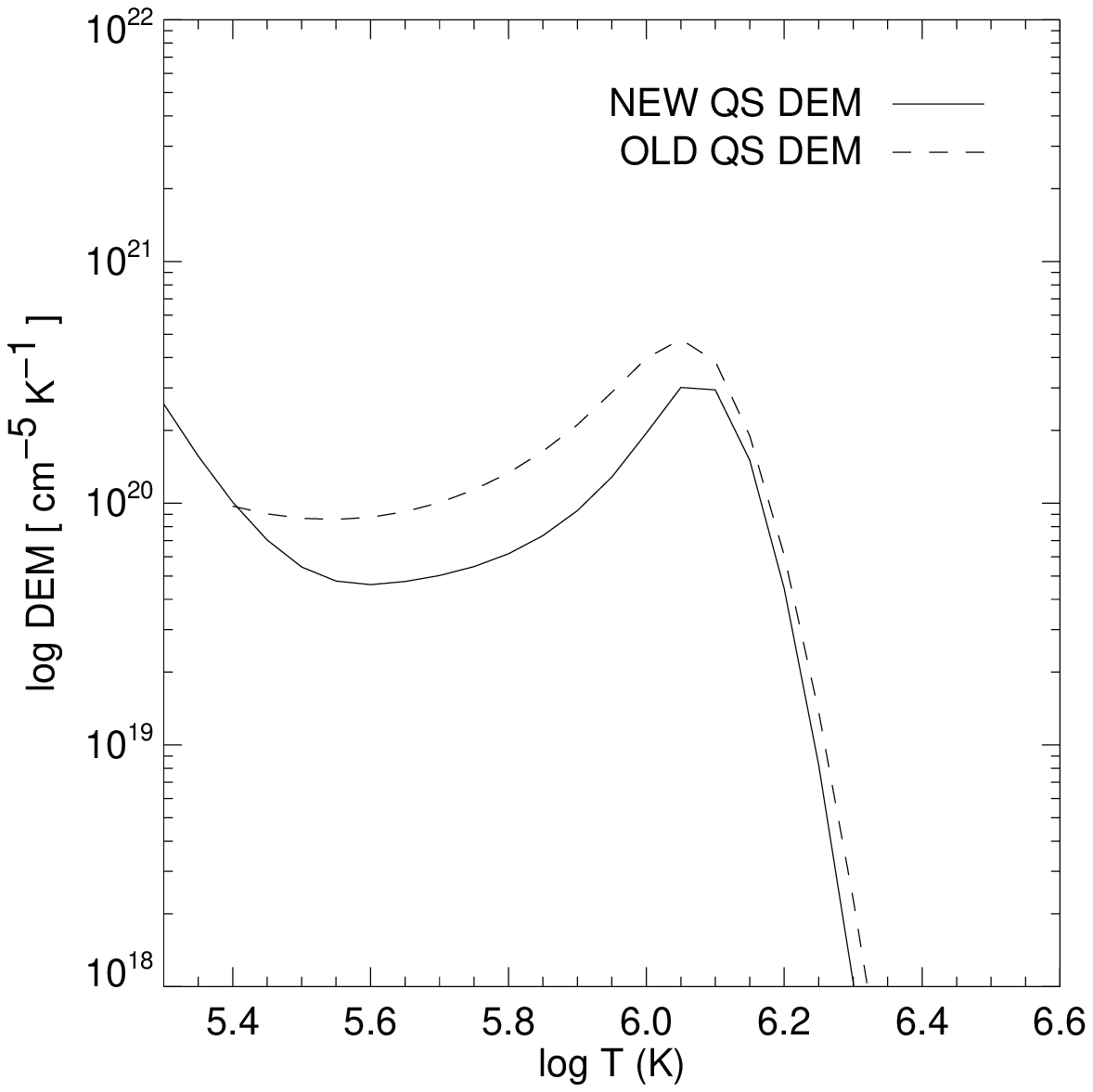}
\includegraphics[width=0.45\linewidth]{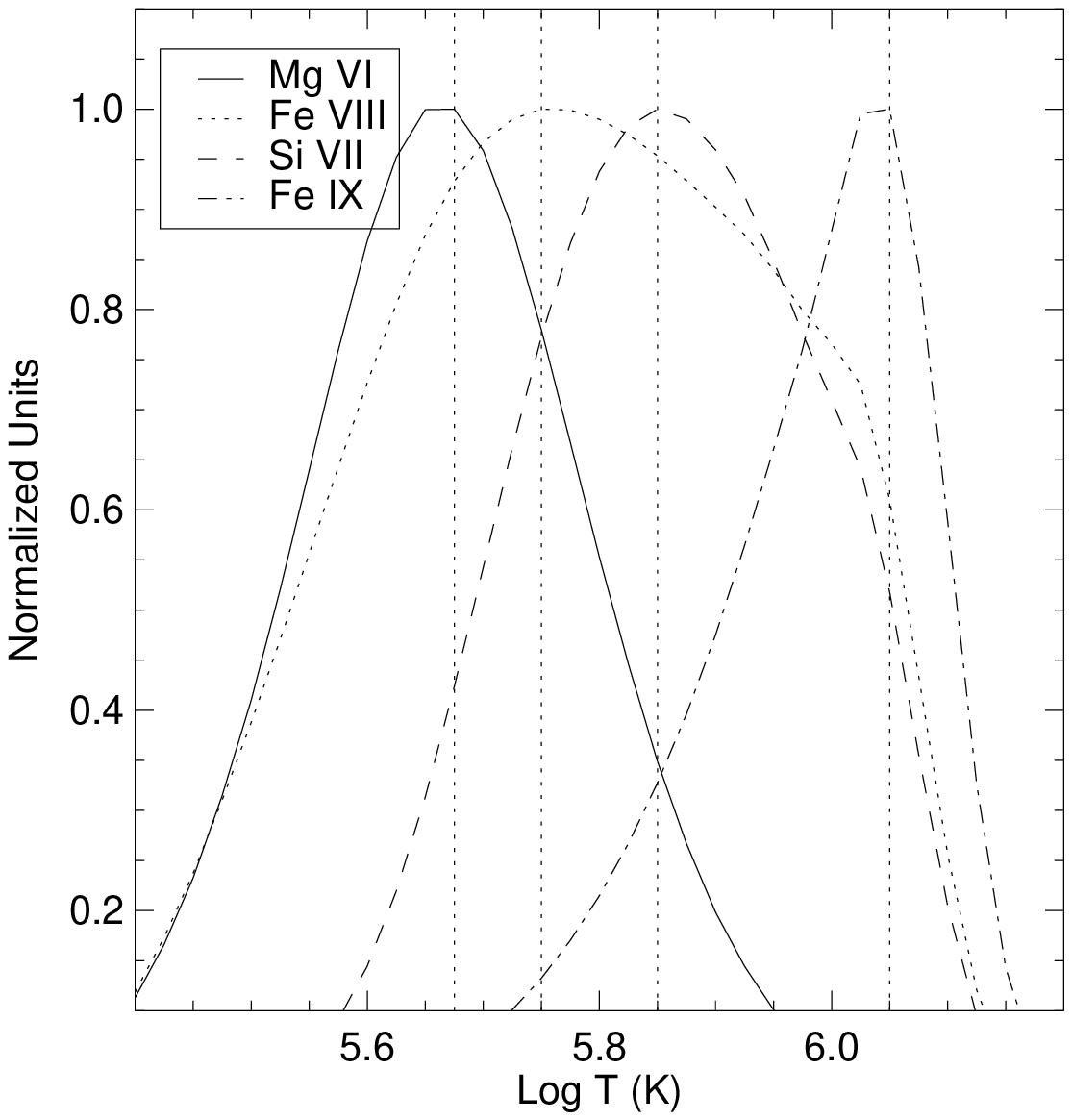}
\caption{Top row: Context observations of the quiet Sun region. From left to right,
images in \ion{Fe}{8} 185.213\,\AA, \ion{Si}{7} 275.368\,\AA, \ion{Fe}{10} 184.536\,\AA, 
\ion{Fe}{11} 188.216\,\AA, \ion{Fe}{12} 195.119\,\AA, \ion{Fe}{13} 202.044\,\AA, 
\ion{Fe}{14} 274.203\,\AA, and \ion{Fe}{15} 284.160\,\AA. The images are
scaled linearly between their minimum and maximum values, except for \ion{Fe}{15}
which is scaled logarithmically because it is very weak.
Images longward of 250\,\AA\, have been shifted upward to account for the offsets between
the EIS CCDs. The common area of the CCDs was averaged to compute the DEM. The Y-dimension
has been rebinned for presentation.
Lower row: Left panel, comparison between the quiet Sun DEM derived using the latest 
atomic data in this paper and that of \citet{brooks_etal2009}.
Right panel, normalized $g (T_e,N_e)$ function in the quiet Sun for 
\ion{Mg}{6} 268.986\,\AA, \ion{Fe}{8} 185.213\,\AA, \ion{Si}{7} 275.368\,\AA, and \ion{Fe}{9} 188.485\,\AA.
The lines are formed at $\log\, (T_e/K)$ = 5.68, 5.75,
5.85, and 6.05, respectively.
\label{fig3}}
\end{figure*}
The function $G$ is strongly peaked at the characteristic temperature of the line, but
the temperature of 
peak contribution to the total intensity can be shifted from the theoretical one if there is relatively
more emitting material at higher or lower temperatures. 
Another key issue for this
paper is that 
if the contribution function is broad, there may be significant contributions to the total intensity from
multiple temperatures. In Figure \ref{fig2}, note that 
the \ion{Fe}{8} curve is significantly broader
than that of \ion{Si}{7} in the new \citet{dere_etal2009} calculation, and encompasses much of the temperature range of that ion. 
A further refinement to the analysis then, is to calculate the
{\it effective formation temperature} (T$_{eff}$) of the line by
convolving the contribution function with the DEM distribution for the target region or structure of interest,
and then taking the temperature at which that function has its maximum value. 
\begin{equation}
g (T_{eff},N_e) = \max ( g (T_e,N_e) )
\label{eq3}
\end{equation}
where $g (T_e,N_e)$ is obtained by
differentiating Equation \ref{eq1}, viz.
\begin{equation}
g (T_e,N_e) = G_{ij} (T_e, N_e) \phi (T_e) 
\label{eq4}
\end{equation}

\subsection{Formation Temperatures in the Quiet Sun}    
\label{qsdem}
Before turning to the AR fan loops, we first investigate the formation temperatures of lines in the
quiet Sun. 
Since we are also interested in how the more recent ionization balance
calculations affect the derived DEM, and whether they could help in resolving known discrepancies, 
we derived a new QS DEM using the 
latest updated atomic data from v6.0.1 of the CHIANTI database and adopting 
the ionization
fractions calculated by \citet{dere_etal2009}. 
Throughout this paper we use the pre-launch laboratory
photometric calibration of EIS \citep{lang_etal2006}.

Perhaps the best way to test the accuracy of the atomic data is to
use off-limb QS observations where there is less contamination along the line of sight from multiple temperatures.
Therefore, we selected lines for analysis that were identified as yielding consistent emission measure distributions
above the quiet solar limb. Most of them are taken from 
Table 4 of \citet{warren&brooks_2009}. The critical issue here, however, is the consistency of lines in the 
temperature range where \ion{Fe}{8}, \ion{Fe}{9}, and \ion{Si}{7} are formed, and it would be best if
we had independent constraints on the DEM that do not involve these lines. Therefore, we used deep exposure
observations of an on-disk QS dataset. 
Deep exposure data are helpful for bringing out the weaker cool lines below $\log\, (T_e/K)$ = 5.8, that 
would be harder to detect off-limb, 
and therefore for constraining the DEM at the lower temperature end.
This also allows us to 
directly compare the results to that of \citet{brooks_etal2009} where various inconsistencies in observed
and predicted intensities were found in on-disk data using
the older ionization balance calculations. 

The data used were taken on 2009 June 14 starting at 09:59:42UT.
Context images of the quiet region are shown in Figure \ref{fig3}
including both \ion{Si}{7} 275.368\,\AA\, and \ion{Fe}{8} 185.213\,\AA, which again look
very similar. The EIS study used the 2$''$ slit with 4 min exposures and scanned a FOV of  60$'' \times$ 512$''$
in about 2hrs 10 mins. About 70\% of the wavelength range of the CCDs were read-out and 
telemetered to the ground. 
Since this does not 
cover the full CCD, some of the lines used in \citet{warren&brooks_2009} were not available.
In addition, we included several important lines of interest for this work, i.e.,
\ion{Mg}{6} 268.986\,\AA, \ion{Fe}{8} 185.213\,\AA, and \ion{Si}{7} 275.368\,\AA, and, as discussed, additional
cooler temperature lines to provide constraints on the DEM 
(\ion{O}{4} 279.631\,\AA, \ion{O}{4} 279.933\,\AA, \ion{O}{5} 248.456\,\AA, \ion{Fe}{8} 186.601\,\AA, \ion{Fe}{8} 194.663\,\AA, \ion{Si}{7} 272.641\,\AA,
\ion{Mg}{7} 278.402\,\AA,
and \ion{Mg}{7} 280.737\,\AA). All the lines used are listed in Table \ref{tab1} which also shows
which lines were selected for the analysis of the fan loops in Sections \ref{ftarcfs} and \ref{daea}.

The data were processed using standard procedures
available in SolarSoft. In addition, however, the orbital variation of the line centroids
and spectral line tilt
were estimated by single Gaussian fits to the \ion{Fe}{12} 195.119\,\AA\, line and removed
from the data prior to further analysis. The offsets
in {\it X}- and {\it Y}- between the short and long wavelength CCDs were corrected for so
that the same area was averaged. These offsets were estimated using the
software developed by \citet{young_etal2009}. The data were then averaged over the common area of
the two CCDs and fitted using single or multiple Gaussians as appropriate.

\subsection{Differential Emission Measure}
\label{qsdem_dem}
The method for determination of the DEM distribution is the same as described in \citet{warren_2005}, \citet{brooks&warren_2006},
and \citet{brooks_etal2009}. Briefly, we represent the DEM with a series of spline knots, the magnitudes
of which are found by $\chi^2$ minimization of the differences between the observed and DEM predicted intensities. This
minimization is performed by the SolarSoft routine MPFIT \citep{markwardt_2009}.
The number and initial values of the spline knots can be pre-set and they can be 
interactively manipulated to control the smoothness of the emission measure. This technique has been
found to be useful for representing rapid changes in the DEM.

We measured the electron density using the following three line ratios:
\ion{Fe}{13} 196.525\,\AA /202.044\,\AA, 
\ion{Fe}{13} 202.044\,\AA/203.826\,\AA, and \ion{Fe}{12} 186.880\,\AA/195.119\,\AA. There are 
a number of issues with blending and the accuracy of the atomic data within certain ranges for 
these diagnostic ratios that are discussed in detail by \citet{young_etal2009}. None of these
issues are expected to affect our measurements significantly, however. The ratios give  
densities for the quiet Sun of $\log\, (N_e/cm^{-3})$ = 8.7--9.0,
indicating a pressure
of $\log\, (P_e/cm^{-3} K^{-1})$ = 14.9--15.2. The upper ends of these ranges are from 
\ion{Fe}{12}, which tends to give higher values \citep{young_etal2009}. Since we cannot 
rule out that the possibility that this is a real density change with temperature, we choose to assume that the
electron pressure is constant. One of these assumptions is necessary to cast the 
intensity integral in the form 
of Equation \ref{eq1} and calculate the contribution functions. 
We adopt a value of $\log\, (P_e/cm^{-3} K^{-1})$ = 15.0 in this work and compute the
contribution functions 
for all the emission lines of Table \ref{tab1} using CHIANTI. 

The quiet Sun DEM of \citet{brooks_etal2009} was used to give a first guess for the temperature
knots and spline fit to the QS data. The fit was then adjusted slightly to improve the agreement 
between observed and predicted intensities. The 
results are given in Table \ref{tab1} and shown in comparison with our previous result in
Figure \ref{fig3}. 80\% of the lines are reproduced to within 30\%.
\ion{Fe}{14} 270.519\,\AA\, and \ion{Fe}{15} 284.160\,\AA\, are significantly overestimated but these lines are very
weak in the spectrum and the DEM is falling rapidly in this temperature range. 
\ion{Fe}{13} 200.021\,\AA, \ion{Fe}{14} 264.787\,\AA, and \ion{Fe}{13} 197.434\,\AA\, are underestimated by the DEM but
these results are consistent with our previous work, and the latter is blended with \ion{Fe}{8} 197.362\,\AA\, \citep{brown_etal2008}. 
\ion{Mg}{7} 278.402\,\AA\, is underestimated by about a factor of two, but this discrepancy can be resolved if
the contribution from the \ion{Si}{7} 278.445\,\AA\, blend is accounted for as described in \citet{young_etal2007b}.

All of the \ion{Fe}{8}, \ion{Fe}{9}, and \ion{Si}{7} lines are consistent with each other and 
the predicted intensities are within 30\% of the observed ones.
These results are interesting because the DEM curve derived using the updated atomic data and ionization fractions from CHIANTI v6.0.1
resolves the inconsistencies found using older versions in several 
previous studies of \ion{Fe}{8} and \ion{Si}{7} lines in the quiet Sun off limb, on disk, and in cool active region features
\citep{warren&brooks_2009,brooks_etal2009,landi&young_2009}. 

\begin{figure*}
\centering
\includegraphics[width=0.44\linewidth,viewport=50 70 350 320,clip]{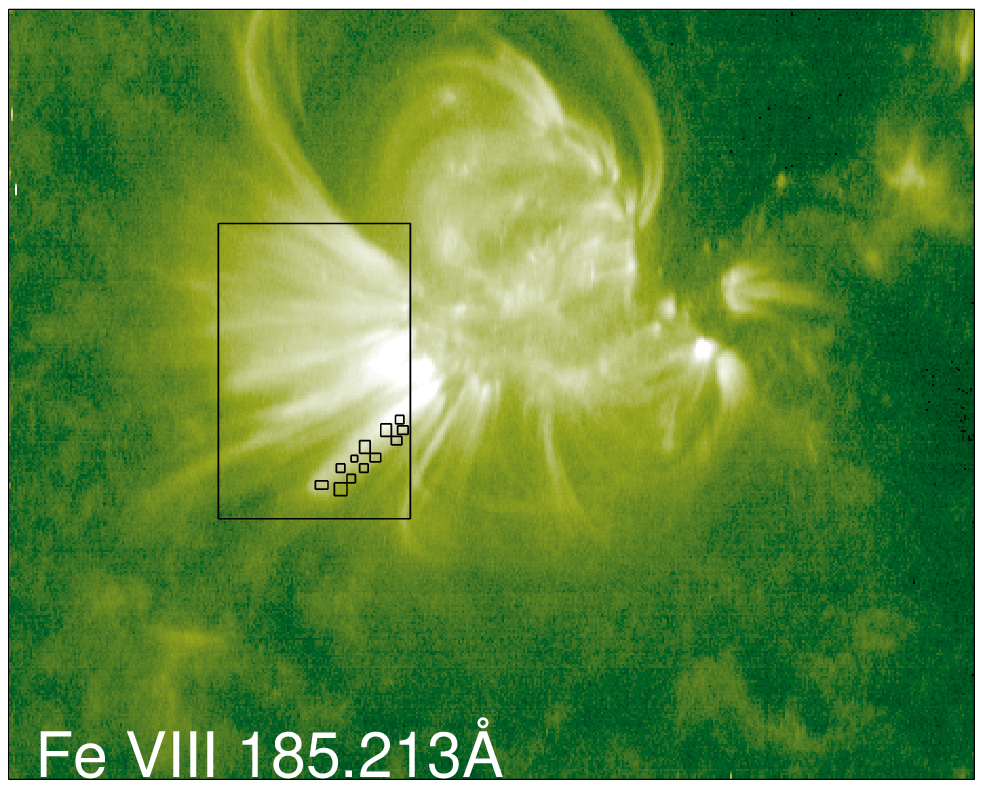}
\includegraphics[width=0.44\linewidth,viewport=50 70 350 320,clip]{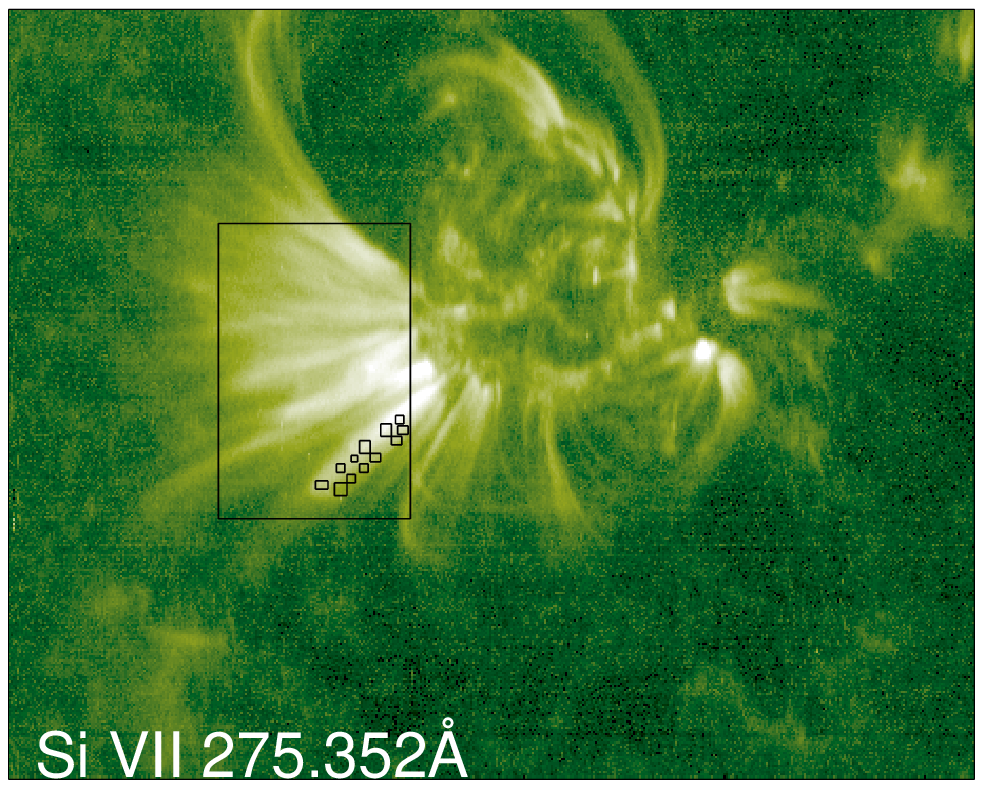}
\includegraphics[width=0.28\linewidth,viewport=60 110 165 270,clip]{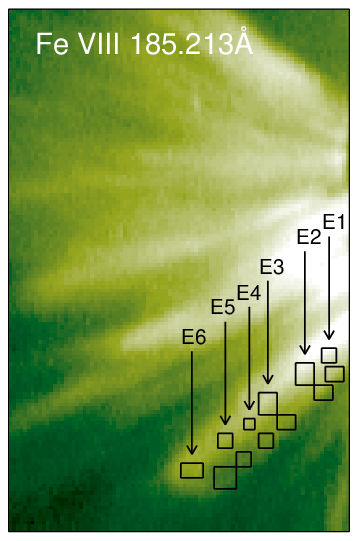}
\includegraphics[width=0.42\linewidth,viewport=50 70 350 340,clip]{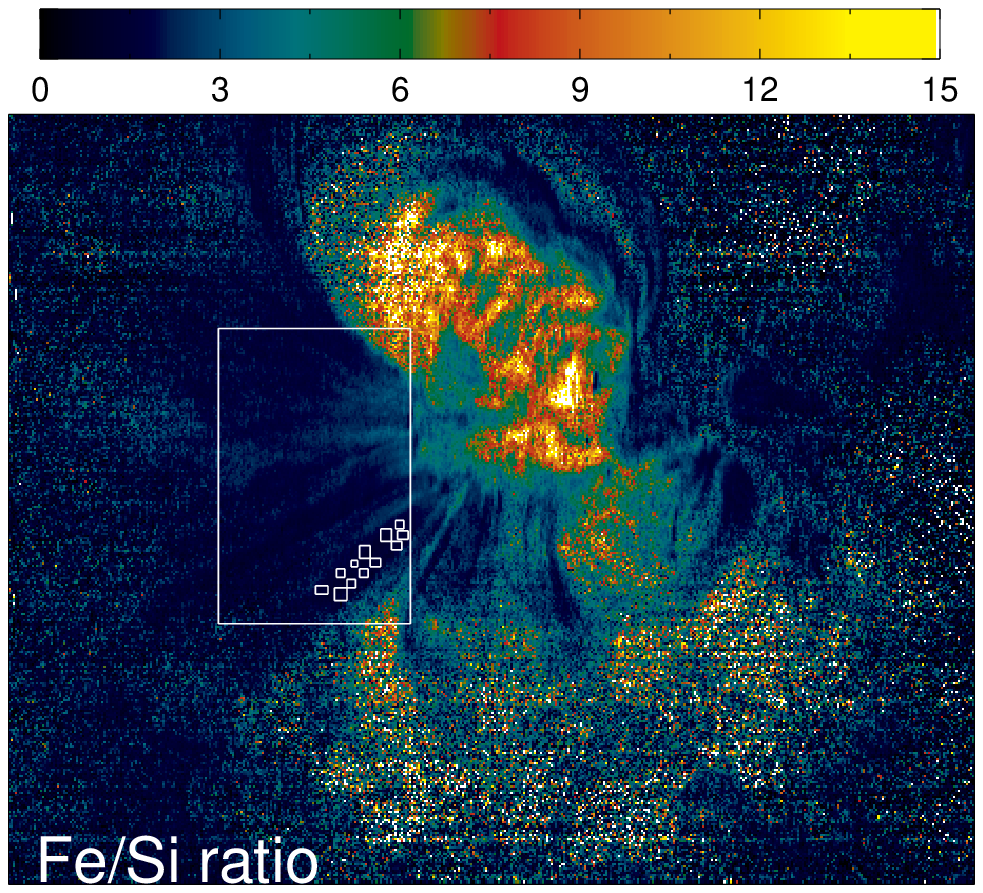}
\includegraphics[width=0.28\linewidth]{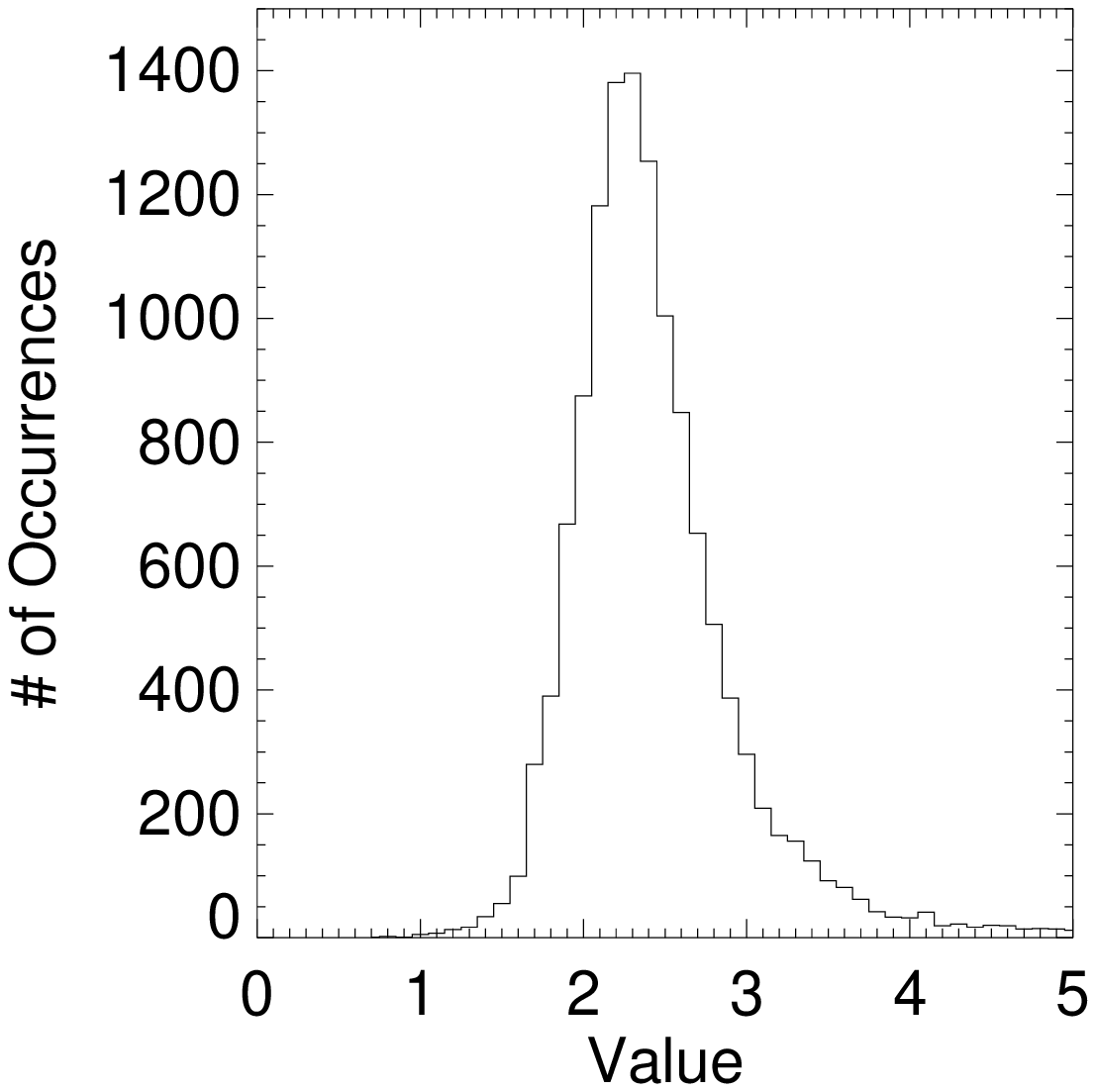}
\caption{Example showing the similarity of \ion{Fe}{8} 185.213\,\AA\, and \ion{Si}{7} 275.368\,\AA\, images.
A ratio map and histogram are shown indicating the low variation of the ratio across the fans. 
The histogram is made from the region indicated by the large box.
An enlargment of the boxed region is also shown to highlight the locations used for the analysis
along the fan loop (E1--E6). 
The adjacent boxes show the areas used for background subtraction.
\label{fig4}}
\end{figure*}

\begin{figure*}
\centering
\includegraphics[width=0.95\linewidth]{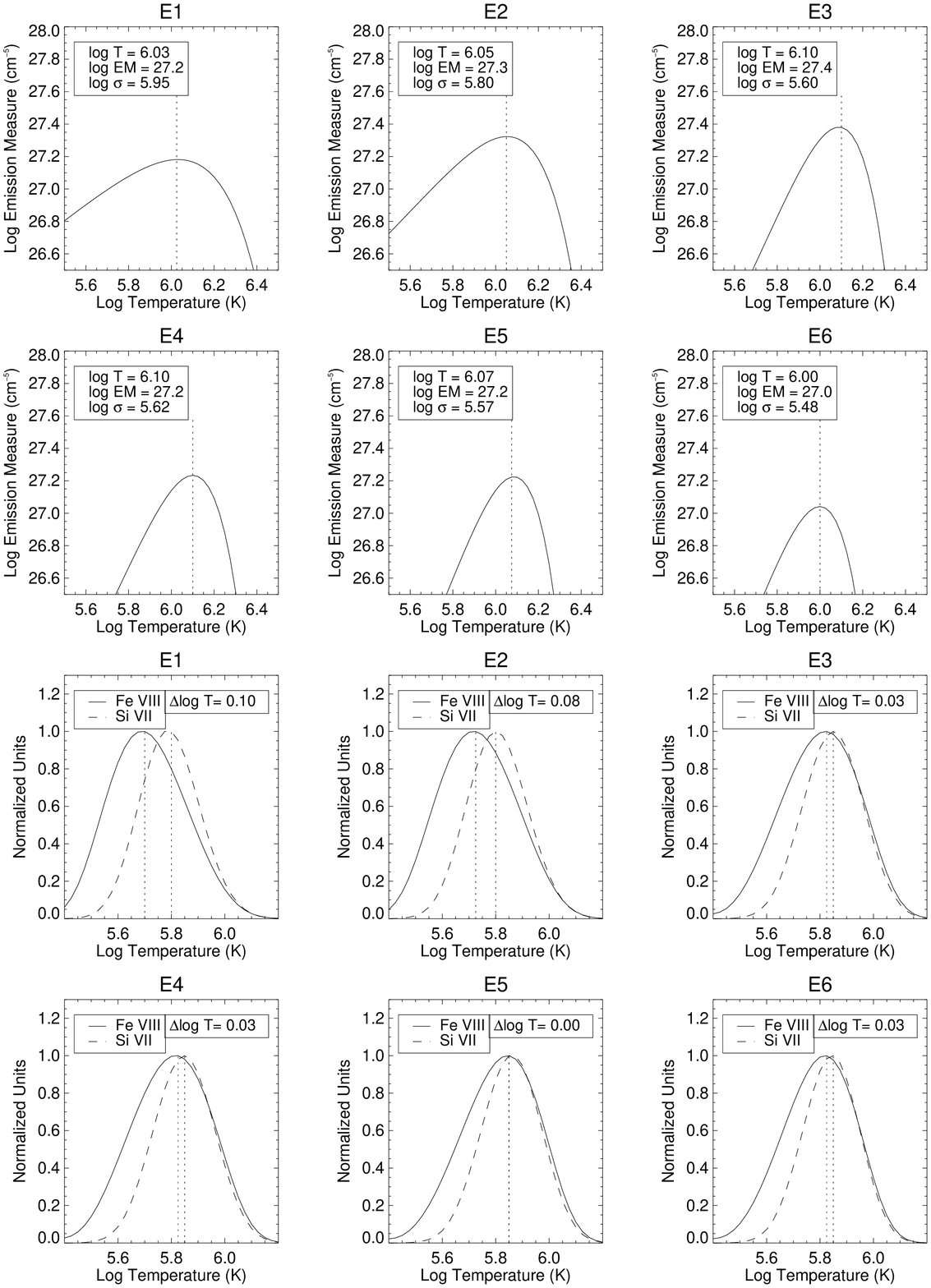}
\caption{Emission measure results along the loop from the base (top left) to the top (bottom right) and the 
effect on the intensities of \ion{Fe}{8} 185.213\,\AA\, and \ion{Si}{7} 275.368\,\AA.
Top two rows: temperature distribution with the peak temperature, emission measure, and Gaussian width indicated. 
Bottom two rows: Intensity of the two
spectral lines as a function of temperature. The vertical dotted lines indicate the formation temperatures ($\log\, T_{max}$). The
separation in formation temperatures is shown in the legend ($\Delta\log T$) and decreases towards the loop top.
\label{fig5}}
\end{figure*}

\subsection{Discussion}
Using Equations \ref{eq3} and \ref{eq4} we determined T$_{eff}$ in the quiet Sun for four key lines
(\ion{Mg}{6} 268.986\,\AA, \ion{Fe}{8} 185.213\,\AA, \ion{Si}{7} 275.368\,\AA, and \ion{Fe}{9} 188.485\,\AA).
The gradient of the DEM slope in the
quiet Sun increases in the temperature range of formation of these lines (Figure \ref{fig3}) so that there 
is relatively more material above the temperature of the maximum ion abundance for all of them. 
This results in all of these lines being formed at temperatures above their maximum ion abundance.
As noted, however, the \ion{Fe}{8} 185.213\,\AA\, line is particulary sensitive to this because of the larger width of
the $F (T_e,N_e)$ curve.
It is therefore disproportionately affected, and is actually formed much
closer in temperature to \ion{Si}{7} 275.368\,\AA\, than the ionization balance would suggest. 
An illustrative example is shown in Figure \ref{fig3}. Note that the $g (T_e,N_e)$ curve for \ion{Fe}{8} 185.213\,\AA\,
is considerably broader than for the other lines, suggesting that there will be contributions from multiple temperatures.
Note also that because the contribution functions for individual lines differ within an ion they may also have different effective
temperatures. 

\section{Formation Temperatures in Active Region Cool Fan Structures}
\label{ftarcfs}
In the previous section we discussed how the slope of the DEM could lead to the \ion{Fe}{8} 185.213\,\AA\,
and \ion{Si}{7} 275.368\,\AA\, lines being formed at similar temperatures in the quiet Sun.
This DEM analysis was made, however, on averaged spectra
over a relatively large area, 
and the differences may not be dramatic because 
the structure of the atmosphere does not change 
rapidly over the small temperature interval ($\log\, (T_e/K)$ = 5.75--5.85) where the 
lines are formed. 
It is clear, however, that specific features such as
the cool fan loops
look very similar in \ion{Si}{7} and \ion{Fe}{8} images. It
remains to be seen whether this explanation could hold for these specific features of interest.

To address this question we investigate the temperature distribution of the extended fan loops
in AR 10978 (shown in Figure \ref{fig1}). This active region has previously been
studied by several authors 
\citep{doschek_etal2008,brooks_etal2008,warren_etal2008a,ugarteurra_etal2009,bryans_etal2010,brooks&warren_2011}.

\subsection{Context images and intensity ratio map}
Context images of the region 
in both \ion{Fe}{8} 185.213\,\AA, and \ion{Si}{7} 275.368\,\AA\, are shown in Figure \ref{fig4}. 
As with the analysis of the QS, we corrected for the CCD offsets,   
orbital variation 
and spectral line tilt
prior to further analysis. In this case we use the artificial neural network model of \citet{kamio_etal2010} to
correct the orbital variation of the line centroids and grating tilt. The rest of the data processing was accomplished
using standard procedures (eis\_prep). The fan region is shown as a large box in Figure \ref{fig4}.

Also shown in Figure \ref{fig4} is a map of the \ion{Fe}{8} 185.213\,\AA/\ion{Si}{7} 275.368\,\AA\, intensity
ratio. This quantitatively describes the similarity of the two images. 
Note that the ratio appears to vary in the active region core.
\ion{Fe}{8} 185.213\,\AA\, is blended with \ion{Ni}{16} 185.251\,\AA\, \citep{young_etal2007a,brown_etal2008},
which is formed at 2.8MK, and it likely contributes to this emission
and influences the ratio. 
In contrast, one can see that the fan region is 
almost uniformly dark, indicating little variation in the ratio across this area. A histogram of the 
ratio values for the pixels within the boxed region is also shown in the Figure. 
A Gaussian fit to this distribution indicates that the mean
value is 2.3 and the standard deviation is 0.34, i.e., the variation in the ratio across the fans is
less than 15\%. 

\subsection{EM analysis}
\label{fan_dem}
We computed the emission measure in several small boxes
extending along one of the fan loops 
(shown as regions E1--E6 in Figure \ref{fig4}).
The intensities were averaged in these areas. The background emission was estimated for  
each box by averaging the intensities in the adjacent boxes also shown in the Figure. 
Single and multiple Gaussian fits
were made as appropriate.

For this analysis we used most of the same lines as we used for the QS. These are indicated
in Table \ref{tab1}. A few lines were not available in the current dataset and could not be
included. These lines can also be identified from the Table. 
Furthermore, we included \ion{O}{6} 183.937\,\AA, \ion{O}{6} 184.117\,\AA, \ion{Fe}{16} 262.984\,\AA, and
\ion{Fe}{16} 265.003\,\AA\, to provide additional low and high temperature constraints. 

\citet{warren_etal2008a} analyzed a sample of `warm' EUV loops in this region by fitting a
Gaussian distribution in temperature for the model emission measure of the form,
\begin{equation}
\xi (T_e) = \frac{EM_0}{\sigma_{T_e}\sqrt{2\pi}}
    \exp\left[-\frac{(T_e-T_0)^2}{2\sigma_{T_e}^2}\right]
\end{equation}
For comparison with this work we use the same fitting technique here.
The fit is made by $\chi^2$ minimization of the differences between the measured and computed intensities.
The density is allowed to be a free parameter in the fitting and we obtain values of $\log\, (N_e/cm^{-3})$ = 9.2--9.7,
decreasing from base to apex. These results are comparable to the decrease along another example reported
by \citet{young_etal2007b}.

The computed temperature distributions 
for positions E1--E6 along the loop together with their peak temperatures, emission measure, 
and Gaussian width are shown in the top two rows of Figure \ref{fig5}. 
There are a few problematic lines that are not well reproduced by the model. They vary by position,
but generally include \ion{O}{4} 279.933\,\AA, \ion{O}{5} 248.456\,\AA, and \ion{O}{6} 183.937\,\AA, all of which
are underestimated to varying degrees. 
Furthermore, the background subtracted high temperature lines of \ion{Fe}{14} and
\ion{Fe}{16} are often underestimated, including the normally robust \ion{Fe}{13} 202.044\,\AA\, line.
\ion{O}{5} 248.456\,\AA\, is blended with an \ion{Al}{8} line
that is a weak contributor in most conditions, hence the agreement found for \ion{O}{5} in the QS. The blend, however, 
could contribute more when the emisison measure is
peaked mostly at coronal temperatures. 
The \ion{Mg}{7} 278.402\,\AA\, and 280.737\,\AA\, lines
also sometimes show unusual behavior. In most cases they are reproduced well, but mismatches of 50--80\% are sometimes
seen. In these cases, the discrepancies for \ion{Mg}{7} 278.402\,\AA\, cannot be resolved by accounting for
the blend with \ion{Si}{7} 278.445\,\AA. 
Since it is mostly the low and high 
temperature lines that show discrepancies it could be that the temperature distribution deviates from a Gaussian
shape in the wings. 
In general most of the lines are reproduced well, however. 
In total, about 80\% of the lines
are reproduced to within 40\% with most better than 30\%. 

\subsection{Discussion}
Compared to the quiet Sun, the fans show a significant enhancement of 
emission measure (factor of 3--7). The distribution peaks at $\log\, (T_e/K)$ = 6.0 near the base (E1) and
increases along the loop to a maximum of $\log\, (T_e/K)$ = 6.1 (E3 and E4) before decreasing again. The 
dispersion in the Gaussian distribution appears to decrease along the loop length. Previous 
studies of cool loops in active regions have suggested that they are isothermal \citep{delzanna&mason_2003}.
Our results indicate that at its narrowest, the width of the emission measure distribution 
is still $\log\, (\sigma_{T_e}/K) \approx$ 5.4. In comparison with the `warm' loops in this region
studied by \citet{warren_etal2008a}, this fan loop has a broader temperature distribution 
along most of its length. Only
towards the tip (E6) does it approach the values for the `warm' loops. 
Note that the gradient of the slope of the emission measure distribution 
in the temperature range $\log\, (T_e/K)$ = 5.6--5.8
is greater than in the quiet Sun, even at the base of the fan (E1). 

\begin{figure*}
\centering
\includegraphics[width=0.44\linewidth,viewport=-15 0 360 360,clip]{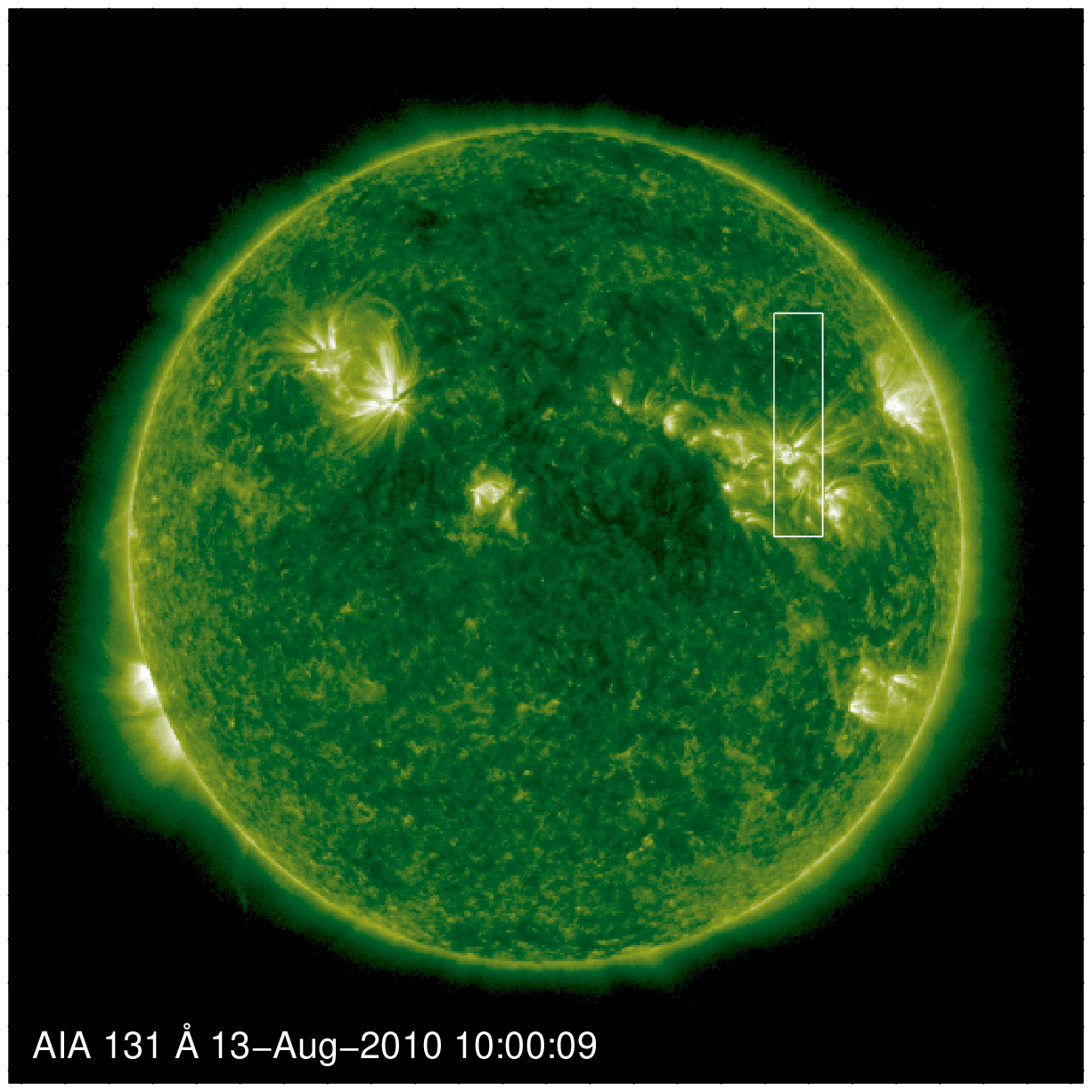}
\includegraphics[width=0.44\linewidth,viewport=-15 0 360 360,clip]{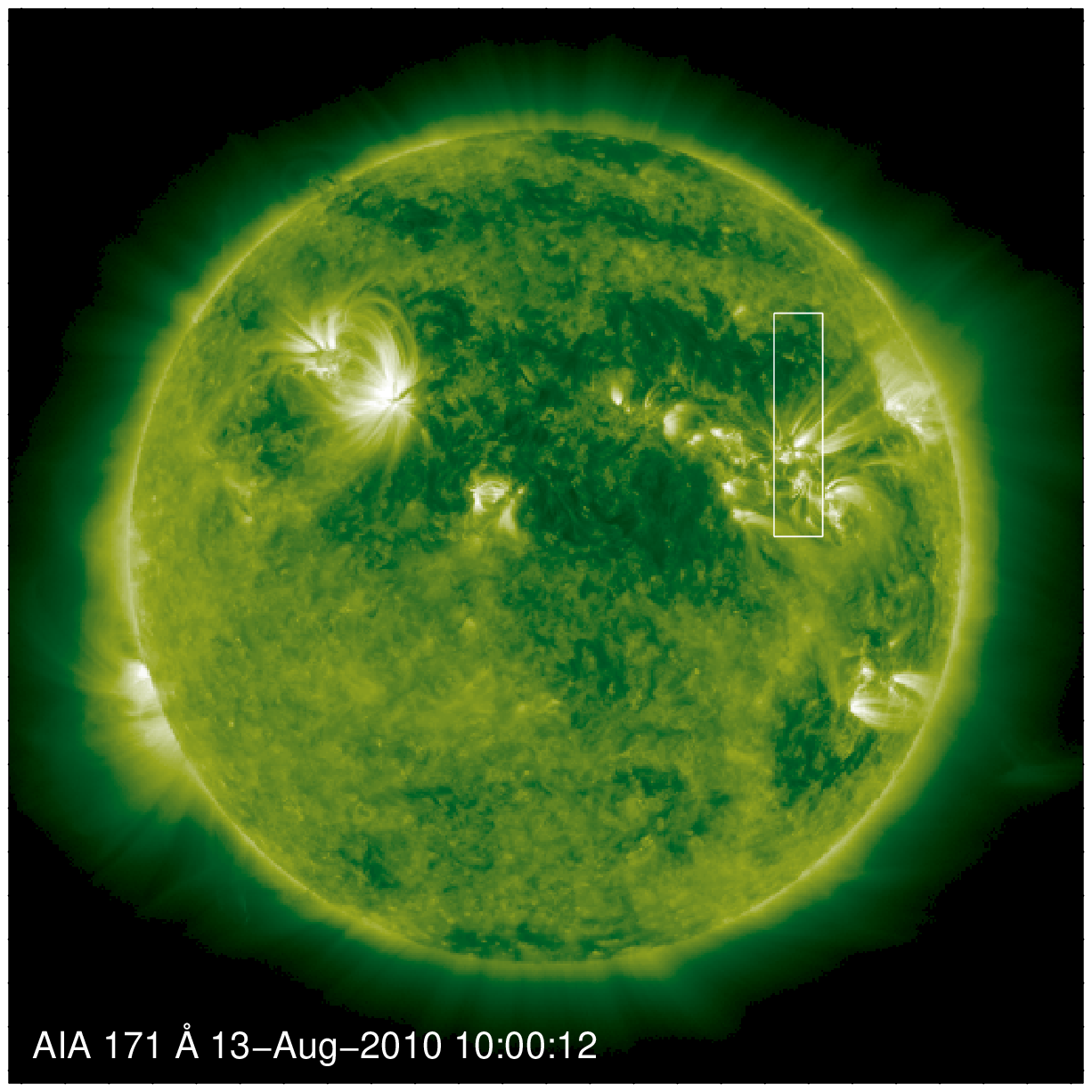}
\includegraphics[width=0.44\linewidth,viewport=-15 0 360 360,clip]{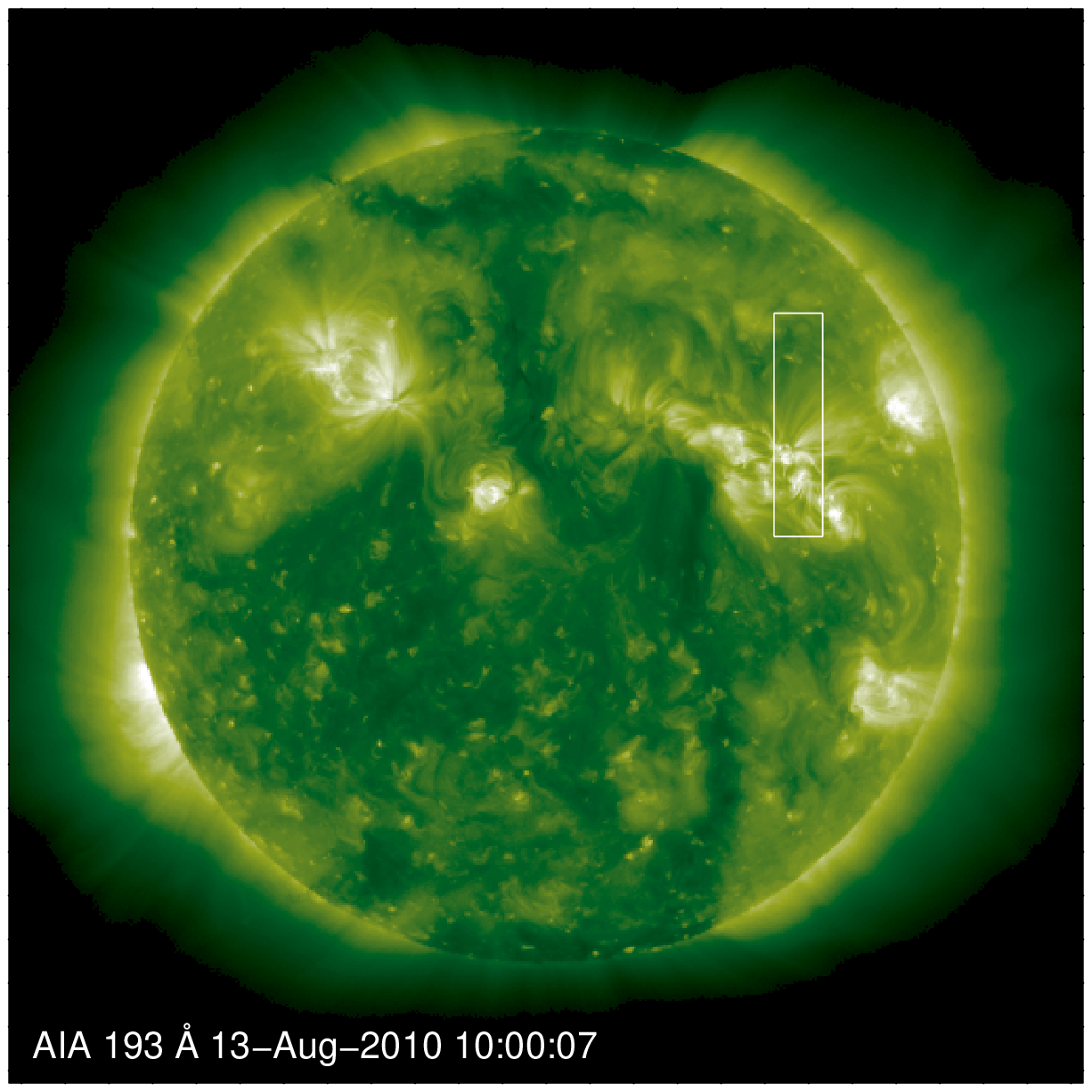}
\includegraphics[width=0.44\linewidth,viewport=-15 0 360 360,clip]{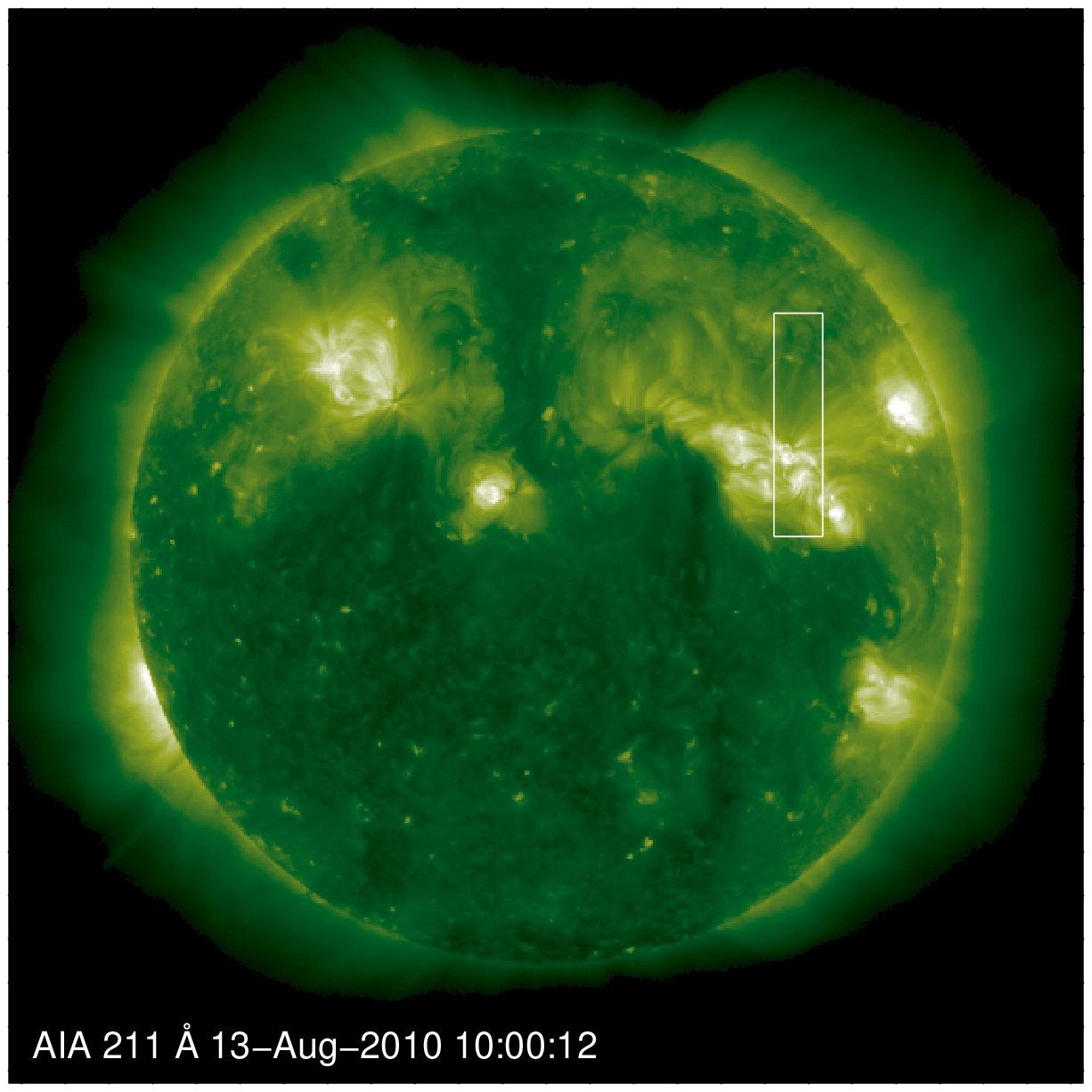}
\caption{Full Sun images from SDO in the AIA 133\,\AA, 171\,\AA, 193\,\AA, and 211\,\AA\, filters. The overlaid box shows the FOV used in
Figure \ref{fig7}.
}
\label{fig6}
\end{figure*}
\begin{figure*}
\centering
\includegraphics[width=1.00\linewidth]{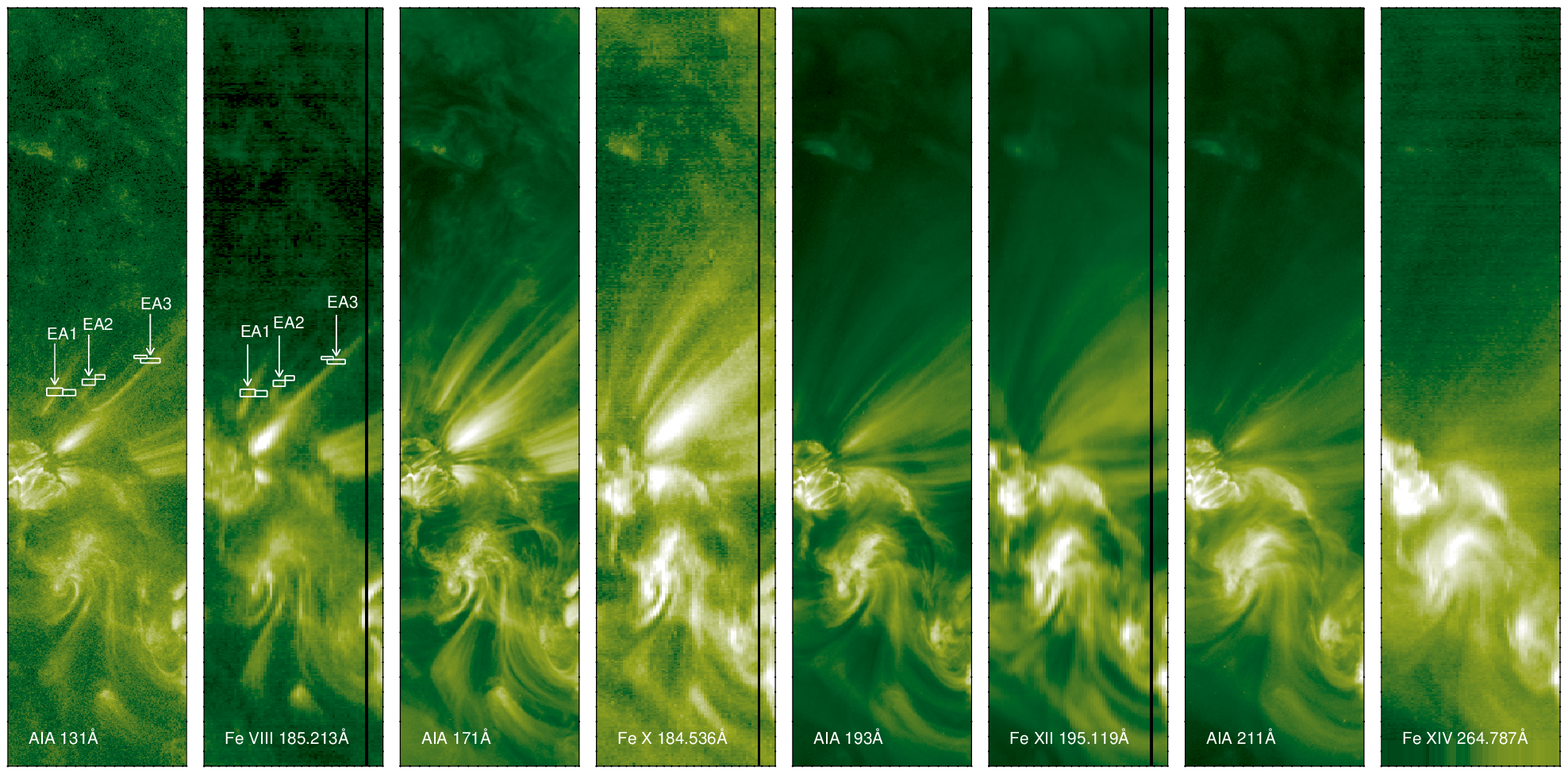}
\caption{AIA and EIS images of fan loops. From the left: 
AIA 133\,\AA, 
EIS \ion{Fe}{8} 185.213\,\AA, 
AIA 171\,\AA, 
EIS \ion{Fe}{10} 184.536\,\AA, 
AIA 193\,\AA, 
EIS \ion{Fe}{12} 195.119\,\AA, 
AIA 211\,\AA\, filters, 
and EIS \ion{Fe}{14} 264.787\,\AA. 
The images are scaled logarithmically.
}
\label{fig7}
\end{figure*}
Figure \ref{fig5} also shows the normalized $g (T_e,N_e)$ function 
for the \ion{Fe}{8} 185.213\,\AA, and \ion{Si}{7} 275.368\,\AA\, lines
that results from convolving their contribution functions with the 
fan emission measure at each position. We see 
that even when the temperature distribution is relatively broad (E1), the effective formation
temperatures of the lines are separated by only 0.1 in the $\log$. It also indicates that the separation
in temperatures reduces as we go along the fan loop until they are both formed at the same temperature
of $\log\, (T_{eff}/K)$ = 5.85. Clearly when the EM distribution is narrow in this 
temperature range the two lines will be formed close together in temperature. This analysis also shows
that this is true even if the EM distribution is relatively broad, provided it has a positive gradient. The 
\ion{Fe}{8} 185.213\,\AA\, line seems to be very sensitive to the slope of the EM distribution in general.

\begin{figure*}
\centering
\includegraphics[width=0.95\linewidth]{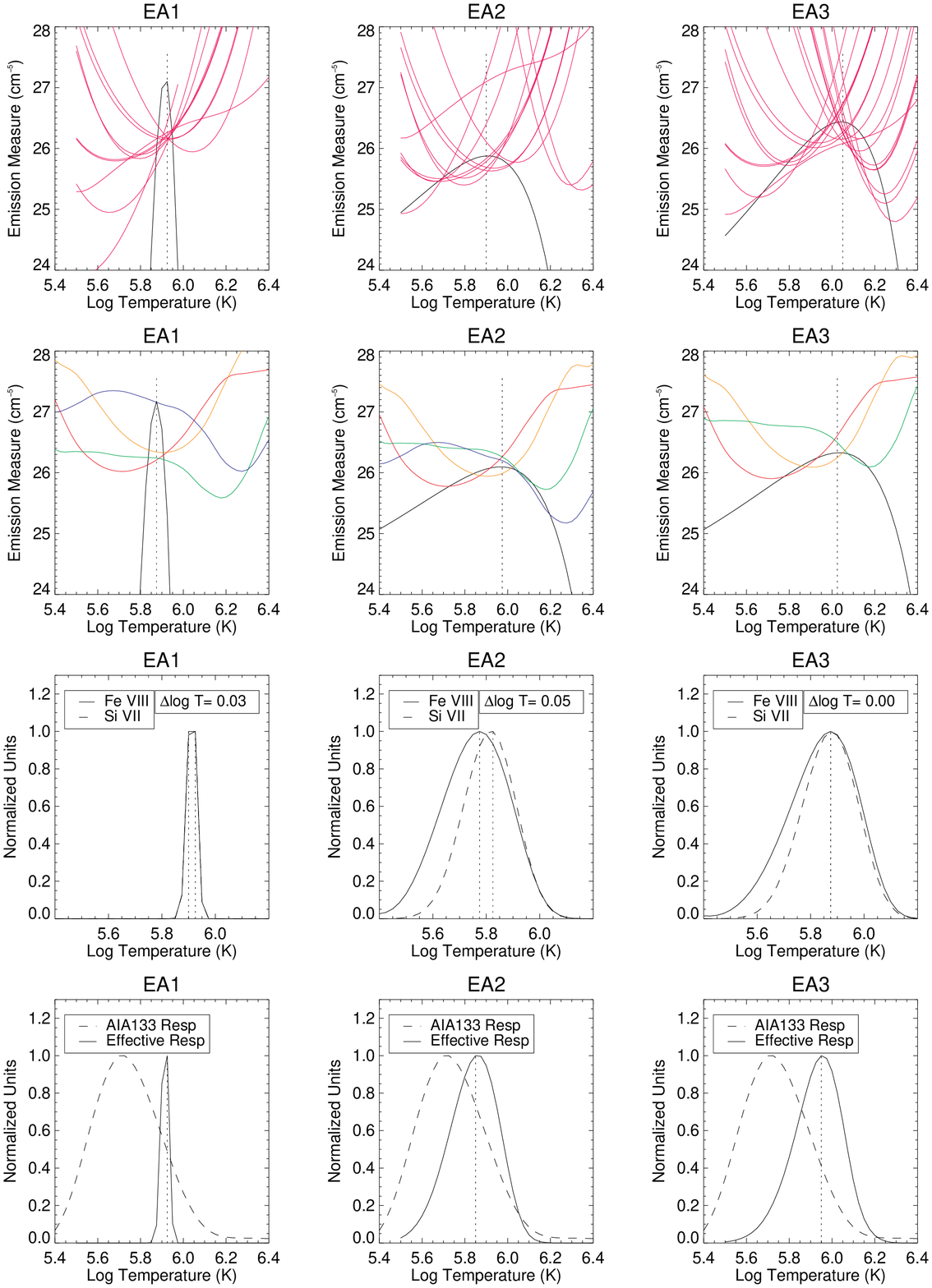}
\caption{Emission measure results for the three loops EA1, EA2, and EA3 from EIS and AIA and the 
effect on the intensities of \ion{Fe}{8} 185.213\,\AA, \ion{Si}{7} 275.368\,\AA, and the AIA 131\,\AA\, response function.
Top row: EIS EM distributions. Second row: AIA EM distributions. 
EM loci curves are overlaid in color. 
Third row: Intensity of the two EIS
spectral lines as a function of temperature. The vertical dotted lines indicate the formation temperatures. The
separation in formation temperatures is shown in the legend.
Bottom row: The AIA 131\,\AA\, effective response as a function of temperature. The original response is shown
by the dashed line.
\label{fig8}}
\end{figure*}
\section{Emission measure analysis with EIS and AIA}
\label{daea}
To examine whether our explanation could hold in general for all fan loops we investigated a number of 
other cases. The loops we selected were observed in active region 11093 on 2010, August 13. 
In this analysis, we also include data obtained by SDO/AIA \citep{golub_2006}.
As mentioned in the introduction, it is also important to understand the formation of spectral lines in detail
if we are to correctly interpret observations by EUV imagers. The broad wavelength pass-bands of such instruments
include multiple lines and their contributions may change in quiet and active conditions. We illustrate this here by 
examining the 
temperature response of the AIA 131\,\AA\, filter. The pass-band of this filter is expected to
contain contributions from \ion{Fe}{8} 130.941\,\AA\, and 131.240\,\AA, \ion{Fe}{20} 132.850\,\AA, \ion{Fe}{21} 128.755\,\AA, and
\ion{Fe}{23} 132.906\,\AA\, with the \ion{Fe}{8} lines the dominant contributor outside of flares. 
As we have seen, the formation of \ion{Fe}{8} lines
is non-trivial. 

Figure \ref{fig6} shows full Sun images from AIA taken at 10UT. AR 11093 was the target for {\it Hinode}
and the EIS FOV is overlaid as a box (120$''$ by 512$''$). The EIS observation started
at 09:26:40 and lasted about 1 hour. During this time the 1$''$ slit was moved in steps of 2$''$ across the region with 60s
exposures at each position. The EIS data were reduced and processed in the same way as discussed previously.

The AIA data were obtained pre-processed from the level 1.5 cutout service. These data have had 
flat-fielding and dark current corrections applied. Bad pixels and cosmic ray hits have also been repaired by an
iterative method. We converted the data to DN/pixel/s. The images have also been rotated to solar North 
and shifted to match the SDO/HMI (Helioseismic and Magnetic Imager) 
FOV center so that they are all coaligned. Nevertheless, we detected some 
small offsets between different filters and corrected this misalignment by cross-correlating the full 
disk images.

Coalignment of the EIS and AIA data was achieved by
extracting the EIS FOV from the AIA 131\,\AA\, full Sun image and cross-correlating it with 
the \ion{Fe}{8} 185.213\,\AA\, EIS raster. Due to the orbital variation of the position of 
the EIS slit, the spatial sampling is not uniform across the raster. In contrast, the AIA
image is taken nearly instantaneously (2.9s exposure) so there is negligible movement. Therefore,
the effective plate-scale magnification between the images is not uniform in the solar X-direction.
For these observations this discrepancy amounted to 1$''$ across half the raster and was corrected manually.

Figure \ref{fig7} shows images from EIS and AIA of the EIS FOV.
For this analysis we selected small boxes on each of 3 loops and averaged the intensities in 
these areas. The selected positions are shown in the figure as EA1, EA2, and EA3. The background emission was estimated
for each box by averaging the emission in the adjacent boxes (also shown in the figure). 
Single and multiple Gaussian fits to the EIS data were made as appropriate. The spectral line-list
for these observations was smaller than that of the December 2007 region, but it still covered
a sufficiently broad temperature range from $\log (T_e/K)$ = 5.5 to 6.4. The selected lines are 
indicated in Table \ref{tab1}. 
In addition, we included \ion{O}{6} 184.117\,\AA,
\ion{Mg}{5} 276.579\,\AA, and \ion{Mg}{6} 270.394\,\AA\,
to provide further lower temperature constraints. 

The emission measure analysis is the same as that used in the previous section for consistency. Since the 
background subtraction sometimes reduces the intensities of the density diagnostic lines
to zero, we have assumed a density of $\log (N_e/cm^{-3})$ = 9.5 in computing
the emission measure for both EIS and AIA. This value is the average of that obtained along the loop analyzed in \S \ref{fan_dem}.

We extracted intensities from the coaligned 131\,\AA, 171\,\AA, 193\,\AA, 
211\,\AA, 335\,\AA, and 94\,\AA\, filters. These filters are expected to be dominated
by lines of \ion{Fe}{8}, \ion{Fe}{9}, \ion{Fe}{12}, \ion{Fe}{14}, \ion{Fe}{16}, and \ion{Fe}{18}.
The 94\,\AA\, filter also contains contributions from \ion{Fe}{8} and \ion{Fe}{10} which can dominate
under certain conditions \citep{odwyer_etal2010}.
We use only the intensities that correlate well with the 131\,\AA\, intensity for generating the emission measure distribution.
In practise, this generally excludes the 335\,\AA\, and 94\,\AA\, bands where there is little emission 
for the fan loops we study.
A photometric calibration uncertainty of 25\% is used as the intensity error (Boerner et al., in preparation).
The response functions were calculated for a grid of electron densities and temperatures spanning 
$\log\, (N_e/cm^{-3})$ = 6--12 and $\log\, (T_e/K)$ = 4--8. This was done by calculating
isothermal spectra as a function of wavelength at each temperature and density, $S(\lambda, T_e, N_e)$, 
using CHIANTI v6.0.1 and the coronal abundances of \citet{feldman_etal1992}. The spectra were then
convolved with the effective areas for each filter, $E_{i} (\lambda)$, and integrated in wavelength,
$\int E_{i} (\lambda) S(\lambda,T_e,N_e) d\lambda$, to obtain the response functions, 
$R_{i} (T_e,N_e)$. The
effective areas were obtained from SolarSoft. 

The EM distributions computed from the EIS data for EA1, EA2, and EA3 are shown in the 
top row of Figure \ref{fig8}. 
The results computed from the AIA data are shown in the second row. EM loci curves are also
shown to indicate the sensitivity of the EIS lines and AIA filters.
Table \ref{tab2} shows the
peak temperature ($\log T_0$), emission measure ($\log EM_0$), and Gaussian width ($\log \sigma_0$) derived from EIS and AIA for each loop.
The uncertainties in these parameters are also given. 

The two instruments 
agree that the peak temperatures of the loops are $\log\, (T_e/K)$ = 5.9--6.0. The emission measure
magnitudes are $\log\, (EM/cm^{-5})$ = 26--27 and the EIS and AIA results are consistent to 25--55\%.
The EIS data indicate that one of the loops (EA1) is isothermal and that the other two have a finite (but narrow) width,
consistent with the findings for the loop in \S \ref{fan_dem}. 
The AIA data yield similar results. For the two loops that are not isothermal, AIA finds a slightly larger 
thermal width (0.14 in $\log$). This may be due to the coarser temperature resolution compared to EIS.
The uncertainties in all parameters from AIA are also larger.
Note that an error of zero is returned
in the isothermal case because the temperature becomes fixed.

Again in Figure \ref{fig8} we show the intensities of \ion{Fe}{8} 185.213\,\AA\, and \ion{Si}{7} 275.368\,\AA\,
as a function of temperature after convolution with the EM distribution of each loop. In all three cases the two
lines are effectively formed at the same temperature. 

We also investigated the effective response of the AIA 131\,\AA\, filter after convolution with 
the EIS emission measure distribution.
The main contributing lines in this pass band at $\log\, (T_e/K) <$ 6.4 are
\ion{Fe}{8} 130.941\,\AA, and \ion{Fe}{8} 131.240\,\AA. These produce the peak
around $\log\, (T_e/K)$ = 5.7 in the usual response, $R_{131} (T_e)$ (dashed line in Figure \ref{fig8}).
The convolved function $R_{131} (T_e) \xi (T_e) T_e$ is plotted in Figure \ref{fig8} as the solid line.
Note that
the peak of the response shifts to higher temperatures after being convolved with the EM distribution; 
$\log\, (T_e/K)$ = 5.85--5.95.
This is consistent with the analysis in the previous sections and demonstrates that 
these effects can also be important
for imager observations.

\section{Summary}
Motivated by the desire to study the temperature structure of active region fan loops we have attempted
to resolve inconsistencies found in previous work using EIS data. In particular,
we have shown that the similarity in EIS \ion{Fe}{8} 185.213\,\AA\, and \ion{Si}{7} 275.368\,\AA\, images,
that is not expected from the respective temperatures of peak abundance in ionization equilibrium, can be
understood when a more accurate calculation of the effective formation temperature in the solar corona is performed.
This is done by convolving the contribution functions with the DEM of the target of interest.
If the DEM has a steep gradient in the $\log\, (T_e/K)$ = 5.6--5.8 range, or is sharply peaked, the
two lines will be formed close in temperature.
In this work, we compared the effect of this technique on the formation temperatures of these lines in 
the quiet Sun. 
The initial separation of $\log\, (T_e/K)$ = 5.6--5.8 is reduced to 5.75--5.85, and it is clear
that \ion{Fe}{8} 185.213\,\AA\, has a substantial contribution from emitting material at $\log\, (T_e/K)$ = 5.8.

To examine whether this explanation could work for the fan loops, 
we derived the EM distribution along one example in AR 10978. The temperature distribution peaks near 1--1.2MK and narrows along
the loop. The peak contribution to the line intensity is $\log\, (T_e/K)$ = 5.9 for both the
\ion{Fe}{8} 185.213\,\AA\, and \ion{Si}{7} 275.368\,\AA\, lines. 
To investigate whether this effect is generally applicable to other AR fans, we examined
a number of other loops. We found that in all cases the two lines are formed at the same temperature
($\log\, T_e/K \sim$ 5.9).
This suggests therefore, that 
the expected difference in images of the fans formed from these lines is a result of an
overestimation of the separation in formation temperatures by the approximate method of assuming
the lines are formed at the temperature of the peak fractional abundance in ionization equilibrium. 

Note that other lines may be affected in similar ways. For example,
in Figure \ref{fig1} the \ion{Mg}{7} and \ion{Fe}{9} images look different despite the fact that
they have similar 
ionization equilibrium temperatures. We have verified that
they are in fact formed at different temperatures in the fan loop
of Section \ref{ftarcfs}.

To demonstrate the importance of understanding the formation of the EUV spectrum for broad pass-band
imagers we studied the effect of convolving the AIA 131\,\AA\, response function 
with our EIS fan loop EM distributions. 
We showed that, as a result, the peak of the dominant lower temperature
part of the effective response function 
shifts up to $\log\, (T_e/K) \sim$ 5.9. 

It is important to emphasize that we have not shown that this explanation holds
in general for all active areas or structures on the Sun. If the DEM slope is shallower (or flat) 
in the $\log\, (T_e/K)$ = 5.6--5.8 range then the
\ion{Fe}{8} 185.213\,\AA\, and \ion{Si}{7} 275.368\,\AA\, lines should still be formed at a wider separation in 
temperatures and examples of significantly different images should be found. The apparent lack of such observations
for any solar feature, however, provides a
stringent constraint on the gradient of the DEM slope in this temperature range all over the Sun. 
This is consistent with other independent studies that show strong similarities in the shape of 
the DEM distribution in different areas of the quiet Sun
\citep{lanzafame_etal2005,brooks_etal2009,feldman_etal2009a}.

The DEM-gradient resolves the inconsistencies in formation temperatures between the \ion{Fe}{8}
and \ion{Si}{7} lines, but a related issue is that discrepancies have also been found in the 
magnitude of the intensities of these lines in previous DEM studies using EIS data. We showed here
that these additional issues are resolved when the most recent
ionization balance compilation data of \citet{dere_etal2009} are used for the atomic calculations. 
From this analysis, therefore, no substantive relative error in the Fe ionization balance is 
indicated for the specific ions of \ion{Fe}{8} and \ion{Fe}{9}. 

This work has demonstrated that the strong \ion{Fe}{8} 185.213\,\AA\, and \ion{Si}{7} 275.368\,\AA\, lines 
can be used with confidence for DEM studies and
velocity work in the transition region. Therefore, we examined the temperature structure of a small
sample of fan loops. We found that they 
have peak temperatures in the range 0.8--1.2MK. One loop was found to be isothermal, but more often the 
temperature distribution has a narrow width. 
This result is similar to that found by \citet{warren_etal2008a} for `warm' active region loops. 
In one detailed case, the EM distribution is found to broaden considerably towards the base. This could have implications
for the location of the heating. 

We also found that the peak temperatures and emission measures derived from AIA data are in agreement with those derived from EIS.
There is also agreement on whether the loops are isothermal or not. The AIA analysis indicates a slightly larger 
thermal width than EIS when the loops are not isothermal. This is possibly because 
the EIS
data contain observations from consecutive ionization stages of \ion{Fe}{0} whereas the AIA 
data only sample every second ionization stage. 

\acknowledgements
We would like to thank the anonymous referee for helping us to focus and clarify the paper.
We also thank Viggo Hansteen and John Mariska for helpful comments, and 
Paul Boerner and Mark Weber for 
providing the AIA effective areas.
This work was performed under contract with the Naval Research Laboratory and was funded by the 
NASA {\it Hinode} program.
{\it Hinode} is a Japanese mission developed and launched by ISAS/JAXA,
with NAOJ as domestic partner and NASA and STFC (UK) as international partners.
It is operated by these agencies in co-operation with ESA and NSC (Norway).
CHIANTI is a collaborative project involving the NRL (USA), RAL (UK), and the following Universities:
College London (UK), Cambridge (UK), George Mason (USA), and Florence (Italy).
The AIA data are courtesy of SDO (NASA) and the AIA consortium. 

{\it Facilities:} \facility{Hinode (EIS)}

\clearpage

\begin{deluxetable}{lcr@{}lccccc}
\tabletypesize{\footnotesize}
\tablewidth{0pt}
\tablecaption{Quiet Sun intensities and line selection.}
\tablehead{
\multicolumn{1}{l}{Ion} &
\multicolumn{1}{c}{$\lambda_{obs}$ (\AA)} &
\multicolumn{2}{c}{I$_{obs}^a$} &
\multicolumn{1}{c}{I$_{dem}^a$} &
\multicolumn{1}{c}{Ratio} &
\multicolumn{1}{c}{E$_{QS}$} &
\multicolumn{1}{c}{E$_{fan}$} &
\multicolumn{1}{c}{EA$_{fan}$}
}
\tablenotetext{}{Intensities are reported for lines used in the QS DEM
analysis. The ticks indicate the lines that were used for the QS DEM (E$_{QS}$), 
and the EIS (E$_{fan}$) and EIS-AIA (EA$_{fan}$) analyses of the fans in 
sections \ref{ftarcfs} and \ref{daea}. }
\tablenotetext{}{Units are erg cm$^{-2}$ s$^{-1}$ sr$^{-1}$.}
\startdata
   \ion{O}{4} & 279.631 &     1.82$\pm$ & 0.40 &    2.03 & 1.12 & $\surd$ &    &    \\
   \ion{O}{4} & 279.933 &     3.90$\pm$ & 0.86 &    4.06 & 1.04 & $\surd$ & $\surd$ &    \\
   \ion{O}{5} & 248.456 &     4.55$\pm$ & 1.00 &    4.23 & 0.93 & $\surd$ & $\surd$ &    \\
  \ion{FE}{8} & 185.213 &    28.55$\pm$ &     6.28 &   35.70 & 1.25 & $\surd$ & $\surd$ & $\surd$ \\
  \ion{FE}{8} & 186.601 &    21.50$\pm$ &     4.73 &   24.37 & 1.13 & $\surd$ & $\surd$ & $\surd$ \\
  \ion{FE}{8} & 194.663 &     7.79$\pm$ & 1.71 &    6.23 & 0.80 & $\surd$ & $\surd$ &    \\
  \ion{Mg}{6} & 268.986 &     1.21$\pm$ & 0.27 &    1.48 & 1.22 & $\surd$ & $\surd$ &    \\
  \ion{Si}{7} & 272.641 &     4.28$\pm$ & 0.94 &    4.48 & 1.05 & $\surd$ &    &    \\
  \ion{Si}{7} & 275.352 &    14.14$\pm$ &     3.11 &   14.58 & 1.03 & $\surd$ & $\surd$ & $\surd$ \\
  \ion{Mg}{7} & 278.402 &    15.81$\pm$ &     3.48 &    9.01 & 0.57 & $\surd$ & $\surd$ & $\surd$ \\
  \ion{Mg}{7} & 280.737 &     2.22$\pm$ & 0.49 &    2.83 & 1.27 & $\surd$ & $\surd$ &    \\
  \ion{FE}{9} & 188.485 &    21.87$\pm$ &     4.81 &   23.66 & 1.08 & $\surd$ & $\surd$ & $\surd$ \\
  \ion{Fe}{9} & 189.941 &    11.38$\pm$ & 2.50 &   14.13 & 1.24 & $\surd$ &    &    \\
  \ion{FE}{9} & 197.858 &    15.15$\pm$ &     3.33 &   15.91 & 1.05 & $\surd$ & $\surd$ & $\surd$ \\
 \ion{FE}{10} & 184.536 &    74.42$\pm$ &    16.37 &   63.20 & 0.85 & $\surd$ & $\surd$ & $\surd$ \\
 \ion{Fe}{11} & 180.401 &   139.19$\pm$ &    30.62 &  169.40 & 1.22 & $\surd$ &    &    \\
 \ion{FE}{11} & 182.167 &    23.31$\pm$ & 5.13 &   29.01 & 1.24 & $\surd$ & $\surd$ &    \\
 \ion{FE}{11} & 188.216 &    75.23$\pm$ &    16.55 &   81.34 & 1.08 & $\surd$ & $\surd$ & $\surd$ \\
 \ion{Fe}{11} & 192.813 &    21.37$\pm$ &     4.70 &   16.99 & 0.80 & $\surd$ & $\surd$ & $\surd$ \\
 \ion{FE}{12} & 186.880 &    14.63$\pm$ &     3.22 &   15.11 & 1.03 & $\surd$ & $\surd$ & $\surd$ \\
 \ion{Fe}{12} & 192.394 &    20.17$\pm$ &     4.44 &   26.27 & 1.30 & $\surd$ & $\surd$ & $\surd$ \\
 \ion{Fe}{12} & 193.509 &    41.03$\pm$ & 9.03 &   55.33 & 1.35 & $\surd$ & $\surd$ &    \\
 \ion{FE}{12} & 195.119 &    63.04$\pm$ &    13.87 &   82.05 & 1.30 & $\surd$ & $\surd$ & $\surd$ \\
 \ion{FE}{12} & 196.640 &     5.62$\pm$ & 1.24 &    4.82 & 0.86 & $\surd$ & $\surd$ &    \\
 \ion{FE}{13} & 196.525 &     0.76$\pm$ & 0.17 &    0.90 & 1.18 & $\surd$ & $\surd$ &    \\
 \ion{Fe}{13} & 197.434 &     4.10$\pm$ & 0.90 &    1.07 & 0.26 & $\surd$ &    &    \\
 \ion{Fe}{13} & 200.021 &     4.15$\pm$ & 0.91 &    3.18 & 0.77 & $\surd$ &    &    \\
 \ion{FE}{13} & 202.044 &    26.08$\pm$ &     5.74 &   21.47 & 0.82 & $\surd$ & $\surd$ & $\surd$ \\
 \ion{FE}{13} & 203.826 &    10.98$\pm$ &     2.42 &   10.15 & 0.92 & $\surd$ & $\surd$ & $\surd$ \\
 \ion{FE}{14} & 264.787 &     4.76$\pm$ &     1.05 &    3.39 & 0.71 & $\surd$ & $\surd$ & $\surd$ \\
 \ion{Fe}{14} & 270.519 &     0.82$\pm$ & 0.18 &    2.14 & 2.61 & $\surd$ &    & $\surd$ \\
 \ion{FE}{14} & 274.203 &     4.78$\pm$ & 1.05 &    4.37 & 0.91 & $\surd$ & $\surd$ &    \\
 \ion{FE}{15} & 284.160 &     1.97$\pm$ &     0.43 &    5.95 & 3.02 & $\surd$ & $\surd$ & $\surd$ \\
  \ion{Mg}{5} & 276.579 &    &    &    &    &    &    & $\surd$ \\
  \ion{Mg}{6} & 270.394 &    &    &    &    &    &    & $\surd$ \\
   \ion{O}{6} & 183.977 &    &    &    &    &    & $\surd$ &    \\
   \ion{O}{6} & 184.117 &    &    &    &    &    & $\surd$ & $\surd$ \\
 \ion{Fe}{16} & 262.984 &    &    &    &    &    & $\surd$ &    \\
 \ion{Fe}{16} & 265.003 &    &    &    &    &    & $\surd$ &    \\
\enddata
\label{tab1}
\end{deluxetable}

\begin{deluxetable}{lr@{$\pm$}lr@{$\pm$}lr@{$\pm$}lr@{$\pm$}lr@{$\pm$}lr@{$\pm$}l}
\tabletypesize{\scriptsize}
\tablewidth{0pt}
\tablecaption{Comparison of DEM results from EIS and AIA.}
\tablehead{
\multicolumn{1}{l}{} &
\multicolumn{6}{c}{EIS} &
\multicolumn{6}{c}{AIA} \\
[.3ex]\cline{2-7}\cline{8-13} \\[-1.6ex] 
\multicolumn{1}{l}{Loop } &
\multicolumn{2}{c}{$\log T_0$} &
\multicolumn{2}{c}{$\log EM_0$} &
\multicolumn{2}{c}{$\log \sigma_0$} &
\multicolumn{2}{c}{$\log T_0$} &
\multicolumn{2}{c}{$\log EM_0$} &
\multicolumn{2}{c}{$\log \sigma_0$} 
}
\startdata
EA  1 & 5.93 & 0.01 & 27.10 & 0.02 & 4.50 & 0.00 & 5.88 & 0.05 & 27.18 & 0.07 & 4.50 & 0.00 \\
EA  2 & 5.90 & 0.04 & 25.88 & 0.09 & 5.41 & 0.18 & 5.97 & 0.13 & 26.10 & 0.10 & 5.55 & 0.20 \\
EA  3 & 6.05 & 0.01 & 26.44 & 0.02 & 5.46 & 0.04 & 6.03 & 0.08 & 26.33 & 0.10 & 5.60 & 0.27 \\
\enddata
\label{tab2}
\end{deluxetable}
\end{document}